\documentclass[12pt]{article}
\usepackage{fullpage}
\usepackage{microtype}
\usepackage{setspace}
\usepackage{atveryend}
\usepackage{lscape}
\usepackage{amsmath}
\usepackage{amsthm}
\usepackage{longtable}
\usepackage{acronym}
\usepackage{graphicx}
\usepackage{amssymb}
\usepackage{bm}
\usepackage[singlelinecheck=false]{caption}
\usepackage[labelsep=period]{caption}
\usepackage{appendix}
\usepackage{mathtools}
\usepackage{booktabs}
\usepackage{adjustbox}
\usepackage{multirow}
\usepackage{amsmath}
\usepackage{color}
\usepackage[colorlinks=true,citecolor=blue,linkcolor=blue]{hyperref}
\usepackage{natbib}
\usepackage{morefloats}
\usepackage{natbib}
\everydisplay{\abovedisplayshortskip\belowdisplayshortskip}%
\makeatletter \oddsidemargin  -.1in \evensidemargin -.1in

\numberwithin{equation}{section}
\usepackage{paralist}
\usepackage{varioref}
\labelformat{equation}{\textup{(#1)}}
\let\orgautoref\autoref
 \renewcommand{\autoref}
        {\def\equationautorefname{Eq.}%
         \def\figureautorefname{Fig.}%
         \def\subfigureautorefname{Fig.}%
         \def\sectionautorefname{Section}%
         \def\subsectionautorefname{Subsection.}%
         \def\subsubsectionautorefname{Subsubsection.}%
         \def\Itemautorefname{item}%
         \def\tableautorefname{Table}%
         \orgautoref}
       \providecommand{\autorefs}
        {\def\equationautorefname{Eqs.}%
         \def\figureautorefname{Figs.}%
         \def\subfigureautorefname{Figs.}%
         \def\sectionautorefname{Sects.}%
         \def\subsectionautorefname{Sects.}%
         \def\subsubsectionautorefname{Sects.}%
         \def\Itemautorefname{items}%
         \def\tableautorefname{Tables}%
         \orgautoref}

\title{Shrinkage Estimation and Prediction for Joint Type-II Censored Data from Two Burr-XII Populations}

\author{Soheila Akbari Bargoshad\footnote{Email: s.akbari@tabrizu.ac.ir} 
and Hossein Bevrani\footnote{Corresponding author, email: bevrani@gmail.com}\\
\small { Faculty of Mathematics, Statistics and Computer Science, University of Tabriz, Tabriz, Iran}\\
}
\date{}

\begin{document}
\maketitle
\bibliographystyle{agsm}
\setcounter{page}{1} \setcounter{equation}{0} \setcounter{figure}{0}
\setcounter{table}{0}

\noindent{\bf { Abstract:}} 
The main objective of this paper is to apply linear and pretest shrinkage estimation techniques to estimating the parameters of two 2-parameter Burr-XII distributions. Furthermore, predictions for future observations are made using both classical and Bayesian methods within a joint type-II censoring scheme. The efficiency of shrinkage estimates is compared to maximum likelihood and Bayesian estimates obtained through the expectation-maximization algorithm and importance sampling method, as developed by \cite{Akbari Bargoshadi2023} in "Statistical inference under joint type-II censoring data from two Burr-XII populations" published in Communications in Statistics-Simulation and Computation". For Bayesian estimations, both informative and non-informative prior distributions are considered. Additionally, various loss functions including squared error, linear-exponential, and generalized entropy are taken into account.  Approximate confidence, credible, and highest probability density intervals are calculated. To evaluate the performance of the estimation methods, a Monte Carlo simulation study is conducted. Additionally, two real datasets are utilized to illustrate the proposed methods.

\vspace{0.2cm}



\noindent{{} }

\section{Introduction}\label{sec 1}
In survival analysis, censoring refers to the phenomenon of having only partial information on the failure time of a unit due to the study design or data limitation. There are different types of censoring, and the most important one being the type-II censoring scheme. Type-II censoring occurs when an event of interest is observed until a predetermined number of events occurs in the study. Once the specified number of events have occurred, the remaining individuals or components are censored. There are scenarios where it becomes necessary to compare two populations derived from separate lines under identical environmental conditions. For example, in the study conducted by \cite{Balakrishnan and Rasouli2008} and \cite{Rasouli and Balakrishnan2010}, the failure times of air conditioning systems from two Boeing 720 jet airplanes were examined.  In this scenario, two independent samples are taken from each population, with sizes denoted as $m$ and $n$, and subjected to a life-testing experiment simultaneously. The experimenter decides to terminate the experiment after a certain number of failures, denoted as $r$, occur, where $1\leq r \leq N$. Here, $N$ represents the total sample size. This censoring scheme allows for a comparison between the two populations based on the observed failures and is known as a joint type-II censoring scheme.  

The Burr-XII distribution is a continuous probability distribution widely used in statistical modeling and analysis. It is known for its flexibility in accommodating a wide range of shapes, making it suitable for various applications such as lifetime data modeling, survival, and extreme value analyses. The Burr-XII distribution is applied in fields such as reliability engineering, actuarial science, and finance, where it can model variables like insurance claims, stock returns, and time-to-failure data. Notably, the Burr-XII distribution includes several well-known distributions as special cases, including the exponential, gamma, Weibull, and log-logistic distributions.

Previous research has extensively analyzed the Burr-XII distribution in the context of censored and jointly censored data. For example, \cite{Ghitany2002} considered maximum likelihood estimation (MLE) of the parameters for the Burr-XII distribution based on randomly right censored data and presented the necessary and sufficient conditions that determine the existence and uniqueness of the MLEs. \cite{Mousa and Jaheen2002} and \cite{Mousa2002} studied Bayesian and  MLEs  for the parameters and reliability function of the Burr-XII distribution and Baysian prediction censored data using two samples prediction techniques under progressive type-II censored samples, respectively. \cite{Soliman2005} discussed the estimation of lifetime parameters (reliability and hazard functions) and Burr-XII model parameters using MLE and Bayesian methods with progressively type-II censored samples. \cite{Wu2007} focused on the estimation problems associated with the Burr-XII distribution using progressive type II censoring with random removals, utilizing the discrete uniform distribution to model the number of units removed at each failure time. \cite{Li2007} studied empirical estimators of reliability performances for the Burr-XII distribution using progressively type-II censored samples and the linear-exponential (LINEX) loss function. \cite{Abdel-Hamid2009} examined constant-partially accelerated life tests using progressively type-II censored samples, for items with Burr-XII distribution lifetimes, deriving likelihood equations and MLEs , as well as approximate confidence intervals (ACIs) to measure uncertainty around the estimated parameters. \cite{Lee et al2009} applied data transformation to construct an MLE of $C_L$ (where C represents the process capability index and L is the lower specification limit) for the Burr- XII distribution based on progressively type-II right censored samples. \cite{Wu et al2010} considered the estimation of a two-parameter Burr-XII distribution using type-II progressive censoring, proposing pivotal quantities for confidence intervals and regions, and using minimum confidence length and area criteria for optimal estimation. \cite{Wang and Cheng2010} studied presents an MLE via the expectation-maximization (EM) algorithm for estimating Burr-XII parameters with multiple censored data, including methods for constructing confidence intervals and computing asymptotic variance and covariance. E-Bayesian method to estimate the parameters and reliability function of the Burr-XII distribution using type-II censored samples and based on squared error (SE) and LINEX loss functions were considered by \cite{Jaheen and Okasha2011}. \cite{Lio and Tsai2012} studied  the MLE of $\delta=p(X<Y)$ for Burr-XII distribution under the progressively first failure-censored samples and constructed confidence intervals of $\delta$ using two bootstrapping procedures and an asymptotic distribution of the MLE of $\delta$. The MLEs of Burr-XII parameters in constant-stress partially accelerated life tests using two methods, observed-data likelihood function and complete-data likelihood function under multiply censored data, were considered by \cite{Cheng and Wang2012}. \cite{Soliman et al.2013} discussed a new life test plan called a progressively first-failure-censoring scheme to obtain MLE and Bayes estimate for some survival time parameters like reliability and hazard functions, as well as the parameters of the Burr-XII distribution. \cite{Rastogi and Tripathi2013} used hybrid censored data to infer parameters of a two parameter Burr-XII distribution. MLEs using the EM algorithm, ACIs and Bayesian estimates and the corresponding highest posterior density (HPD) intervals were obtained. Statistical inference and prediction on Burr-XII parameters using type-II censored samples were discussed by \cite{Panahi and Sayyareh2014}. Also, \cite{Panahi and Sayyareh2016} focused on the statistical inference and prediction of the Burr-XII distribution using unified hybrid-censored data and developed MLEs employing the EM algorithm and Bayesian estimates of the unknown parameters using approximation methods, namely Lindley's approximation and the Markov chain monte carlo technique. \cite{Kayal et al.2017} estimated unknown parameters of a Burr-XII distribution under type-I hybrid censored data, using the EM algorithm, Lindley method, and Metropolis-Hastings algorithm. Also obtained predictive estimates of censored observations and constructed ACI and prediction intervals. The reliability function of Burr- XII distribution was inferred using a generalized variable method with random removals based on progressively type-II censoring with random removals by \cite{Gunasekera2018}. The Bayesian inference for the parameters of a randomly censored Burr-XII distribution with proportional hazard functions was presented by \cite{Danish et al.2018}. \cite{Okasha et al.2020} developed E-Bayesian predictive functions based on observed order statistics and two samples, under type-II censoring scheme from the Burr-XII model. \cite{Kohansal2020} obtained the estimates of $R=P(X<Y)$ when the variables $X$ and $Y$ follow Burr-XII distributions with distinct parameters under a hybrid progressive censoring schemes.  Using an adaptive type-II progressive censoring scheme, the parameters, reliability, and hazard functions of the Burr-XII distribution were estimated by \cite{Yan et al.2021}. Under a type-I joint competing risks samples, the MLE,  Bayesian estimator, ACI, bootstrap confidence, and Bayesian credible intervals (CrIs) were discussed for Burr-XII distribution by \cite{Abushal et al.2021}. \cite{Ragab et al.2021} estimated the unknown parameters for two Burr-XII distributions under a joint type-I generalized hybrid censoring scheme and using MLE and the Bayes method. \cite{Saini et al.2022} explored classical and Bayesian estimation of multicomponent stress-strength reliability for the Burr-XII distribution using progressively first-failure censored samples.  \cite{Cetinkaya2023} considered the Burr-XII distribution and applied a generalized progressive hybrid censoring scheme to estimate the probability of $R = P(X > Y)$. 

Despite extensive research on the Burr-XII distribution, there has been a lack of investigation into linear shrinkage (LS) and shrinkage pretest (SP) estimators for this distribution under a joint type-II censoring scheme. Moreover, classical and Bayesian prediction methods for this distribution have not been applied under such a scheme. This research gap motivates this paper, which aims to achieve two primary goals: (1) computing LS and SP estimates of the unknown parameters using both classical and Bayesian methods, and comparing them and (2) predicting future failures using classical and Bayesian approaches and determining classical and Bayesian prediction intervals. This study considers three distinct forms of loss functions: the SE,  LINEX loss functions, and the generalized entropy (GE) loss function. The SE loss function is defined as $L_{SE}(\widehat{\theta},\theta)=(\widehat{\theta}-\theta)^2$, where $\widehat{\theta}$ represents an estimator of $\theta$. This loss function penalizes both overestimation and underestimation equally, making it a symmetric function, therefore, the LINEX and the GE loss functions are suggested as alternatives. The LINEX loss function, introduced by \cite{Varian1975}, is an asymmetric and convex function defined as $L_{LINEX}(\widehat{\theta},\theta)=e^{v(\widehat{\theta}-\theta)}-v(\widehat{\theta}-\theta)-1$, $v\neq 0$, where $v$ is the shape parameter that determines the direction and magnitude of underestimation ($v > 0$) or overestimation ($v < 0$). The Bayesian estimator of $\theta$ based on the LINEX loss function is given by  $\widehat{\theta} _{LINEX}=- \dfrac{1}{v}\ln \Big(E(e^{- v\theta})\Big)$. The GE loss function, proposed by \cite{Calabria and Pulcini1994}, is defined as $L_{GE}(\widehat{\theta},\theta)=\Big(\dfrac{\widehat\theta}{\theta}\Big)^{k}-k\ln\Big(\dfrac{\hat{\theta}}{\theta}\Big)-1$, $k\neq 0$, where $k$ is the shape parameter that determines the degree of asymmetry in the loss function. Positive values of $k$ emphasize overestimation, while negative values emphasize underestimation. The Bayesian estimator of $\theta$ based on the GE loss function is defined as $\widehat{\theta}_{GE}=\Big(E(\theta^{-k})\Big)^{-1/k}.$

The structure of the paper is as follows: In \autoref{sec2}, the joint type-II censoring scheme for two Burr-XII distributions, and its parameters Bayesian estimation are reviewed based on the research work of \cite{Akbari Bargoshadi2023}. In \autoref{sec3}, LS and SP estimations are presented and compared them to the other estimations in terms of relative efficiency (RE). \autoref{sec4} focuses on predicting the first and second censored data using the best unbiased predictor (BUP) and Bayesian prediction methods. Also, classical and Bayesian confidence intervals are calculated. A comparison of the proposed method based on simulation study and analysis of two real datasets are conducted in \autoref{sec5} and \autoref{sec6}, respectively. Finally, \autoref{sec7} provides a summary and conclusion of the paper.

\section{Model Specification and estimations }\label{sec2}
Suppose that $\textbf{X}=({X_1}, \ldots ,{X_m})^\prime$ and $\textbf{Y}=({Y_1}, \ldots ,{Y_n})^\prime$ are random vectors representing random samples from the first and second population, respectively, where both populations follow a Burr-XII distribution, with the probability density function $f_i$ and survival function $\bar{F}_i$ $(i=1,2)$, respectively, in the following form
 \begin{eqnarray}\label{pdf1:BURR}
&&{f_1}(x;{\theta_1},{\theta_2}) =\theta_1\theta_2 x^{(\theta_2-1)}(1+x^{\theta_2})^{-(\theta_1+1)},\quad x > 0,\quad {\theta_1} > 0,\quad {\theta_2} > 0,\\\nonumber\\
\label{SF1:BURR} &&\bar{F}_1(x;{\theta_1},{\theta_2}) =(1+x^{\theta_2})^{-\theta_1},\quad x > 0,\quad {\theta_1} > 0,\quad {\theta_2} > 0,
\end{eqnarray}
\begin{eqnarray}\label{pdf2:BURR}
&&{f_2}(y;{\theta_3},{\theta_4}) =\theta_3\theta_4 y^{(\theta_4-1)}(1+y^{\theta_4})^{-(\theta_3+1)},\quad y > 0,\quad {\theta_3} > 0,\quad {\theta_4} > 0,\\\nonumber\\
\label{SF2:BURR} &&\bar{F}_2(y;{\theta_3},{\theta_4}) =(1+y^{\theta_4})^{-\theta_3},\quad y > 0,\quad {\theta_3} > 0,\quad {\theta_4} > 0.
\end{eqnarray}

The parameters $\theta_i$ for $i=1,2,3,4$ represent the positive shape parameters of the distributions. Let $N = m + n$ represent the combined lifetime of selected samples from two populations and $r$ $(1\leq r \leq N)$ denotes the total number of observed or failed units. The order statistics of the $N$ units are represented as $W_1< \ldots < W_N$. Under the joint type-II censoring scheme, we present the observed data as $(\textbf{W}, \textbf{S})$, where $\textbf{W} = (W_1, ..., W_r)^\prime$ and $\textbf{S} = (S_1, ..., S_r)^\prime$. In this representation, $S_i$ takes the value $1$ if $W_i$ corresponds to $X$-failures, and $S_i$ is set to $0$ if $W_i$ corresponds to $Y$- failures, where $(i = 1, 2, ..., r)$. Let $M_r = \sum_{i=1}^r S_i$ denote the number of $X$-failures among $(W_1,...,W_r)^\prime$ and $N_r = \sum_{i=1}^r (1 - S_i) = r - M_r$ denote the number of $Y$-failures among $(W_1,...,W_r)^\prime$. The likelihood function for the observed data based on the joint density function of $(\textbf{\textbf{W}},\textbf{S})$ is
\begin{align}
\label{likelihood:data}
L(\boldsymbol\theta;\textbf{w},\textbf{s})=&\dfrac{m!n!}{(m-m_r)!(n-n_r)!}\prod_{i=1}^{r}[f_1(w_i;\theta_1,\theta_2)]^{s_i}[f_2(w_i;\theta_3,\theta_4)]^{1-s_i}\nonumber\\
&\times[\bar{F}_1(w_r;\theta_1,\theta_2)]^{m-m_r}[\bar{F}_2(w_r;\theta_3,\theta_4)]^{n-n_r},
\end{align}
where $\boldsymbol\theta=(\theta_1,\theta_2,\theta_3,\theta_4)^\prime$ and $0 < {w_1}... < {w_r} < \infty$. \cite{Akbari Bargoshadi2023} acquired the MLE and Bayesian estimations of the parameters ${\theta_i}$ by employing the EM algorithm and the importance sampling method, respectively. They also computed the ACIs and CrIs for the unknown parameters. From a Bayesian viewpoint, they considered the prior distributions of $\theta_1\thicksim G(a_1,b_1)$, $\theta_2\thicksim G(c_1,d_1)$, $\theta_3\thicksim G(a_2,b_2)$, $\theta_4\thicksim G(c_2,d_2)$ in which $G(\alpha,\beta)$ represents the gamma distribution. Then, the joint posterior density function of the parameters was given by
\begin{align}
\label{importance:informative}
\pi(\boldsymbol{\theta}|\textbf{w},\textbf{s})\propto &G_{\theta_2}(m_r+c_1,L_1)\times G_{\theta_1|\theta_2}(m_r+a_1,R_1)
\times G_{\theta_4}(n_r+c_2,L_2)\times G_{\theta_3|\theta_4}(n_r+a_2,R_2)\nonumber\\
&\times\dfrac{exp\Bigg\{-\sum\limits_{i=1}^{r}{s_i\ln(1+w_i^{\theta_2})}-\sum\limits_{i=1}^{r}{(1-s_i)\ln(1+w_i^{\theta_4})}\Bigg\}}{R_{1}^{m_r+a_1}R_{2}^{n_r+a_2}},
\end{align}
where 
\begin{flalign*}
\begin{array}{l}
L_1 = d_1- \sum\limits_{i=1}^{r}{s_i\ln w_i},\quad R_1 = b_1 + (m - m_r)\ln(1 + w_r^{\theta_2}) + \sum\limits_{i = 1}^r {s_i\ln (1+w_i^{\theta_2})},\\
L_2 = d_2 - \sum\limits_{i = 1}^{r} {(1 - s_i)\ln w_i},\quad R_2 = b_2 + (n - n_r)\ln(1 + w_r^{\theta_4}) + \sum\limits_{i = 1}^r {(1-s_i)\ln(1+w_i^{\theta_4})},\\
\end{array}
\end{flalign*}
and Bayesian estimations for the parameters ${\theta_i}$ $(i=1,2,3,4)$ based on the SE, LINEX, and GE loss functions were as follows
\begin{equation*}
\hat{\theta}_{i.SE}=\dfrac{\sum\limits_{j=1}^D{\theta_{ij}\Delta\big(\theta_{2j},\theta_{4j}\big)}}{\sum\limits_{j=1}^D{\Delta\big(\theta_{2j},\theta_{4j}\big)}},\quad\quad\hat{\theta}_{i.LINEX}=-\dfrac{1}{v}\ln\Bigg(\dfrac{\sum\limits_{j=1}^D{e^{-v\theta_{ij}}\Delta\big(\theta_{2j},\theta_{4j}\big)}}{\sum\limits_{j=1}^D{\Delta\big(\theta_{2j},\theta_{4j}\big)}}\Bigg),
\end{equation*}
\begin{equation*}
\hat{\theta}_{i.GE}=\Bigg(\dfrac{\sum\limits_{j=1}^D{(\theta_{ij})^{-k}\Delta\big(\theta_{2j},\theta_{4j}\big)}}{\sum\limits_{j=1}^D{\Delta\big(\theta_{2j},\theta_{4j}\big)}}\Bigg)^{-\dfrac{1}{k}},
\end{equation*}
where
\begin{equation}\label{Delta}
\Delta\big(\theta_{2j},\theta_{4j}\big)=\dfrac{exp\Bigg\{-\sum\limits_{i=1}^{r}{s_i\ln(1+w_i^{\theta_{2j}})}-\sum\limits_{i=1}^{r}{(1-s_i)\ln(1+w_i^{\theta_{4j}})}\Bigg\}}{R_{1}^{m_r+a_1}R_{2}^{n_r+a_2}},
\end{equation}
 and the parameters $\theta_{2j}$, $\theta_{1j}$, $\theta_{4j}$, and $\theta_{3j}$ $(j=1,2,...,D)$ are randomly generated $D$ times from distributions $G_{\theta_{2}}(m_r+c_1,L_1)$, $G_{\theta_{1}|\theta_{2}}(m_r+a_1,R_1)$, $G_{\theta_{4}}(n_r+c_2,L_2)$, and      $G_{\theta_{3}|\theta_{4}}(n_r+a_2,R_2)$, respectively. \cite{Soliman2000} presented a class of non-informative (NIN) prior distributions called quasi-density prior, characterized by $\pi(\theta_i) = \frac{1}{\theta_i^\gamma}$, where $\gamma>0$ and $\theta_i>0$, $i=1,2,3,4$. This prior family encompasses several commonly used distributions and as special cases, we consider including Jeffreys prior when $\gamma=1$. 

\section{Shrinkage Estimator}\label{sec3}
The MLE and Bayesian estimation methods are typically based on the available information within the sample, using standard techniques to estimate unknown parameters.  However, in some cases, the researcher may have additional information about the unknown parameter in the form of prior knowledge or an educated guess. This additional information is referred to as non-sample information.  To incorporate non-sample information into the estimation process, shrinkage estimators have been introduced. These estimators improve the accuracy of the estimation by incorporating both sources of information. In this section, the focus is on employing the LS and SP estimation strategy, which is simple to implement and does not require optimization or hyperparameters.

\subsection{Linear shrinkage Estimator}\label{sub sec 1}
The LS estimator $\widehat{\theta}_{i}^{LS}$ of $\theta_{i}$ is a linear combination of the predetermined point $\theta_{i.0}$ and the MLE or Bayesian estimate of  the parameters $\theta_{i}$, and is defined as (see \cite{Ahmed2014})
\begin{equation}\label{LS}
\widehat{\theta}_{i}^{LS}=w\theta_{i.0}+(1-w)\widehat{\theta}_{i},\quad i=1,2,3,4.
\end{equation}

The constant $w$ represents the shrinkage intensity and $w\in[0,1]$. The value of $w$ is determined by the researcher based on a subjective estimation of the prior information or chosen to minimize the mean squared error of the estimator.  When $w= 0$, then the LS estimator reduces to the $\widehat{\theta}_i$ while if $w=1$, it becomes the $\theta_{i.0}$. Now, based on the \autoref{LS}, the LS estimators of the $\theta_i$ can be obtained using the MLE derived from the EM method and the Bayesian estimators, respectively,
\begin{equation*}\label{LS.mle}
\widehat{\theta}_{i.MLE}^{LS}=w\theta_{i.0}+(1-w)\widehat{\theta}_{i.MLE},\quad i=1,2,3,4,
\end{equation*}
and 
\begin{equation*}\label{LS.bayes}
\widehat{\theta}_{i.Bayes}^{LS}=w\theta_{i.0}+(1-w)\widehat{\theta}_{i.Bayes}, \quad i=1,2,3,4.
\end{equation*}
 
\subsection{Shrinkage pretest estimator}\label{sub sec 2}
 \cite{Ahmed2014} proposed a SP estimator strategy as follows
 \begin{equation}\label{sp1}
\widehat{\theta}_{i}^{SP}= \widehat{\theta}_{i}- w(\widehat{\theta}_{i}-\theta_{i.0})I(\ell_i<\chi_{1,\alpha}^{2}),\quad\quad 0\leq w\leq 1, i=1,2,3,4,
\end{equation}
 or $\widehat{\theta}_{i}^{SP}$ can be computed using $\widehat{\theta}_{i}^{LS}$ as
 \begin{equation}\label{sp2}
\widehat{\theta}_{i}^{SP}=\widehat{\theta}_{i}- (\widehat{\theta}_{i}-\widehat{\theta}_{i}^{LS})I(\ell_i<\chi_{1,\alpha}^{2}), \quad\quad\quad\quad\quad\quad \quad i=1,2,3,4,
\end{equation}
where $I(.)$ is an indicator function, $w$ represents the degree of shrinkage, and $\theta _{i.0}$ refers to a predetermined point related to the parameter $\theta_i$. Also $\ell _i$ is the test statistic for the null hypothesis ${H_0}:\theta_i= \theta _{i.0}$ against ${H_1}:\theta\neq \theta _{i.0}$ and define as $\ell _i=\dfrac{r(\widehat{\theta}_i-\theta_{i.0})^2}{var(\widehat{\theta}_i)}$. The ${H_0}$ is rejected when $\ell _i\geq \chi _{1,\alpha }^2$, where $\alpha$ represents a type-I error and $\chi _{1,\alpha }^2 $ indicate the upper $100\% \alpha$ critical value of the ${\chi ^2}$- distribution with 1 degree of freedom. For obtaining SP estimators of the $\theta_i$ based on the MLE derived from the EM method and Bayesian estimators, respectively, is sufficient in \autoref{sp1} put $\widehat{\theta} _{i.MLE}$ and $\widehat{\theta} _{i.Bayes}$ instead of $\widehat{\theta} _{i}$.  

\section{Prediction}\label{sec4}
The main object of this section is to predict future failure lifetimes, denoted as $W_t=W_{r+j}$, for $j = 1, 2, ..., N-r.$ This prediction is based on the joint type-II censored data, represented as $(\textbf{W},\textbf{S})$. To achieve this object, we begin by calculating the joint density function of $\textbf{W}, \textbf{S}, W_{t}$ and $S_{t}$. Subsequently, we determine the conditional density function of $S_{t}$ and $W_{t}$ given the observed data $(\textbf{W}$, $\textbf{S})$. Let $M_t=\sum\limits_{i=1}^{t}S_i$ and $N_t=\sum\limits_{i=1}^{t}(1-S_i)$ represent the cumulative sum of  $X$-failures and  $Y$-failures, respectively, among $(W_1,W_2,...,W_{t})$. There are three cases for the future failure lifetime $W_{t}$:

 \begin{enumerate}
\item[\textbf{ 1:}]
If $m_r< m$ and $n_r =n$, the future failure $W_t$ is certainly an $X$-failure, with $S_t=1$.
\item[\textbf{ 2:}]
If $m_r = m$ and $n_r < n$, the future failure $W_t$ is definitely a $Y$-failure, with $S_t=0$.
 \item[\textbf{ 3:}]
 If $m_r < m$ and $n_r < n$, the future failure $W_t$ can be either an $X$-failure or a $Y$-failure. 
 \end{enumerate}
 
Then, the joint density function for $(W_{t},S_{t},\textbf{W},\textbf{S})$ is given by
\begin{eqnarray}
\label{pdf:w_{r+j}}
f(w_{t},s_{t},\textbf{w},\textbf{s}) = 
\begin{cases}
{f_1}(w_{t},s_{t},\textbf{w},\textbf{s}),\quad{m_r} < m,\quad{n_r} = n,\\
{f_2}(w_{t},s_{t},\textbf{w},\textbf{s}),\quad{m_r} = m,\quad{n_r} < n,\\
{f_3}(w_{t},s_{t},\textbf{w},\textbf{s}),\quad{m_r} < m,\quad{n_r} < n,
 \end{cases}
\end{eqnarray}
where
\begin{align}
f_1(w_t, s_t,\textbf{w},\textbf{s})=&\dfrac{m!n!f_1(w_t;\theta_1,\theta_2)}{(t-r-1)!(m-m_r-t+r)!}\prod_{i=1}^{r}\big[f_1(w_i;\theta_1,\theta_2)\big]^{s_i}\big[f_2(w_i;\theta_3,\theta_4)\big]^{1-s_i}\nonumber\\
&\times\big[\bar{F}_1(w_r;\theta_1,\theta_2)-\bar{F}_1(w_t;\theta_1,\theta_2)\big]^{t-r-1}\big[\bar{F}_1(w_t;\theta_1,\theta_2)\big]^{m-m_r-t+r},
\end{align}

\begin{align}
f_2(w_t, s_t,\textbf{w},\textbf{s})=&\frac{m!n!f_2(w_t;\theta_3,\theta_4)}{(t-r-1)!(n-n_r-t+r)!}\prod_{i=1}^{r}\big[f_1(w_i;\theta_1,\theta_2)\big]^{s_i}\big[f_2(w_i;\theta_3,\theta_4)\big]^{1-s_i}\nonumber\\ 
&\times\big[\bar{F}_2(w_r;\theta_3,\theta_4)-\bar{F}_2(w_t;\theta_3,\theta_4)\big]^{t-r-1}\big[\bar{F}_2(w_t;\theta_3,\theta_4)\big]^{n-n_r-t+r},
\end{align}
and
\begin{align}
f_3(w_t, s_t,\textbf{w},\textbf{s})=&\sum_{s_{r+1}=0}^{1}\cdots\sum_{s_{t}=0}^{1}\dfrac{m!n!\big[f_1(w_t;\theta_1,\theta_2)\big]^{s_t}\big[f_2(w_t;\theta_3,\theta_4)\big]^{1-s_t}}{(m_{t-1}-m_r)!(n_{t-1}-n_r)!(m-m_t)!(n-n_t)!}\prod_{i=1}^{r}\big[f_1(w_i;\theta_1,\theta_2)\big]^{s_i}\nonumber\\
&\times\big[f_2(w_i;\theta_3,\theta_4)\big]^{1-s_i}\big[\bar{F}_1(w_r;\theta_1,\theta_2)-\bar{F}_1(w_t;\theta_1,\theta_2)\big]^{m_{t-1}-m_r}\big[\bar{F}_1(w_t;\theta_1,\theta_2)\big]^{m-m_t}\nonumber\\
&\times\big[\bar{F}_2(w_r;\theta_3,\theta_4)-\bar{F}_2(w_t;\theta_3,\theta_4)\big]^{n_{t-1}-n_r}
\big[\bar{F}_2(w_t;\theta_3,\theta_4)\big]^{n-n_t}.
\end{align}

So, using \autorefs{likelihood:data} and \ref{pdf:w_{r+j}}, the conditional density function of $(W_{t},S_{t})$ given $(\textbf{w,s})$ becomes
\begin{eqnarray}
\label{cpdf:w_{r+j}}
f(w_{t},s_{t}|\textbf{w},\textbf{s}) = 
\begin{cases}
{f_1}(w_{t},s_{t}|\textbf{w},\textbf{s}),\quad{m_r} < m,\quad{n_r} = n,\\
{f_2}(w_{t},s_{t}|\textbf{w},\textbf{s}),\quad{m_r} = m,\quad{n_r} < n,\\
{f_3}(w_{t},s_{t}|\textbf{w},\textbf{s}),\quad{m_r} < m,\quad{n_r} < n,
 \end{cases}
\end{eqnarray}
where 
\begin{align}\label{cdf1:w_r+j}
f_1({w_t},s_{t} |\textbf{w},\textbf{s})=&\dfrac{(m-m_{r})!}{(t-r-1)!(m-m_r-t+r)!}\big[\bar{F}_1(w_r;\theta_1,\theta_2)-\bar{F}_1(w_{t};\theta_1,\theta_2)\big]^{t-r-1}\nonumber\\
&\times\big[\bar{F}_1(w_{t};\theta_1,\theta_2)\big]^{m-m_r-t+r}\big[\bar{F}_1(w_r;\theta_1,\theta_2)\big]^{m_r-m}f_1(w_{t};\theta_1,\theta_2),                                          
 \end{align}                                                                                                                                                       
\begin{align}\label{cdf2:w_r+j}
f_2(w_{t},s_{t} |\textbf{w},\textbf{s})=&\dfrac{(n-n_{r})!}{(t-r-1)!(n-n_r-t+r)!}\big[\bar{F}_2(w_r;\theta_3,\theta_4)-\bar{F}_2(w_{t};\theta_3,\theta_4)\big]^{t-r-1}\nonumber\\
&\times\big[\bar{F}_2(w_{t};\theta_3,\theta_4)\big]^{n-n_r-t+r}\big[\bar{F}_2(w_r;\theta_3,\theta_4)\big]^{n_r-n}f_2(w_{t};\theta_3,\theta_4),                                       
 \end{align}   
                                
and
\begin{align}\label{cdf3:w_r+j}
f_3(w_{t},s_{t} |\textbf{w},\textbf{s})=&\sum_{s_{r+1}=0}^{1} \cdots \sum_{s_{t}=0}^{1}\dfrac{(m-m_r)!(n-n_r)!\big[f_1(w_{t};\theta_1,\theta_2)\big]^{s_{t}}\big[f_2(w_{t};\theta_3,\theta_4)\big]^{1-s_{t}}}{(m_{t-1}-m_r)!(n_{t-1}-n_r)!(m-m_{t})!(n-n_{t})!}\nonumber\\
  &\times\big[\bar{F}_1(w_r;\theta_1,\theta_2)-\bar{F}_1(w_{t};\theta_1,\theta_2)\big]^{m_{t-1}-m_r}\big[\bar{F}_1(w_{t},\theta_1,\theta_2)\big]^{m-m_{t}}\nonumber\\
&\times\big [\bar{F}_2(w_r;\theta_3,\theta_4)-\bar{F}_2(w_{t};\theta_3,\theta_4)\big]^{n_{t-1}-n_r}
\big[\bar{F}_2(w_{t};\theta_3,\theta_4)\big]^{n-n_{t}}\nonumber\\
  &\times\big[\bar{F}_1(w_r;\theta_1,\theta_2)\big]^{m_r-m}\big[\bar{F}_2(w_r;\theta_3,\theta_4)\big]^{n_r-n}.
  \end{align}  

Now, by replacing \autorefs{pdf1:BURR}-\ref{SF2:BURR} in \autorefs{cdf1:w_r+j}-\ref{cdf3:w_r+j}, the conditional density function of $(W_t,S_t)$ given $(\textbf{w},\textbf{s})$, for the three cases are given by
\begin{align}\label{CDF1:W_{r+j}}
 f_1(w_{t},s_{t}|\textbf{w},\textbf{s})=&\dfrac{(m-m_r)!}{(t-r-1)!(m-m_r-t+r)!}\Big[\big(1+w_r^{\theta_2}\big)^{-\theta_1}-\big(1+w_t^{\theta_2}\big)^{-\theta_1}\Big]^{t-r-1}\nonumber\\
 &\times\big(1+w_{t}^{\theta_2}\big)^{-\theta_1(m-m_r-t+r+1)-1}\big(1+w_{r}^{\theta_2}\big)^{-\theta_1(m_r-m)}\theta_1\theta_2w_t^{\theta_2-1},
 \end{align}

 \begin{align}\label{CDF2:W_{r+j}}
 f_2(w_t,s_{t}|\textbf{w},\textbf{s})=&\dfrac{(n-n_r)!}{(t-r-1)!(n-n_r-t+r)!}\Big[\big(1+w_r^{\theta_4}\big)^{-\theta_3}-\big(1+w_t^{\theta_4}\big)^{-\theta_3}\Big]^{t-r-1}\nonumber\\
 &\times\big(1+w_{t}^{\theta_4}\big)^{-\theta_3(n-n_r-t+r+1)-1}\big(1+w_{r}^{\theta_4}\big)^{-\theta_3(n_r-n)}\theta_3\theta_4w_t^{\theta_4-1},
 \end{align}
 and
  \begin{align}\label{CDF3:W_{r+j}}
 f_3(w_{t},s_{t}|\textbf{w},\textbf{s})=&\sum_{s_{r+1}=0}^{1} \cdots \sum_{s_{t}=0}^{1}\dfrac{(m-m_r)!(n-n_r)!}{(m_{t-1}-m_r)!(n_{t-1}-n_r)!(m-m_{t})!(n-n_{t})!}\nonumber\\
 &\times\Big[(1+w_{r}^{\theta_2})^{-\theta_1}-(1+w_{t}^{\theta_2})^{-\theta_1}\Big]^{m_{t-1}-m_r}\big(1+w_{t}^{\theta_2}\big)^{-\theta_1(m-m_{t-1})-s_t}\nonumber\\
  &\times\Big[(1+w_{r}^{\theta_4})^{-\theta_3}-(1+w_{t}^{\theta_4})^{-\theta_3}\Big]^{n_{t-1}-n_r}\big(1+w_{t}^{\theta_4}\big)^{-\theta_3(n-n_{t-1})-(1-s_t)}\nonumber\\
  &\times\theta_1^{s_t}\theta_2^{s_t}\theta_3^{1-s_t}\theta_4^{1-s_t} w_{t}^{-s_t(\theta_4-\theta_2)+\theta_4-1}(1+w_{r}^{\theta_2})^{-\theta_1(m_r-m)}(1+w_{r}^{\theta_4})^{-\theta_3(n_r-n)}.
 \end{align}
 
Additionally, using the binomial expansion for the terms $\big[(1+w_r^{\theta_2})^{-\theta_1}-(1+w_t^{\theta_2})^{-\theta_1}\big]^{t-r-1}$ and $\big[(1+w_r^{\theta_4})^{-\theta_3}-(1+w_t^{\theta_4})^{-\theta_3}\big]^{t-r-1}$, we can express \autorefs{CDF1:W_{r+j}}, and \ref{CDF2:W_{r+j}}, respectively, in the following forms

\begin{align}
f_1(w_t,s_t |\textbf{w},\textbf{s})=&\dfrac{(m-m_r)!}{(m-m_r-t+r)!}\sum\limits_{j_1=0}^{t-r-1}c_{j_1}(t-r-1)\big(1+w_{t}^{\theta_2}\big)^{-\theta_1(m-m_r-t+r+j_1+1)-1}\nonumber\\
&\big(1+w_{r}^{\theta_2}\big)^{-\theta_1(m_r-m+t-r-j_1-1)}\theta_1\theta_2 w_{t}^{\theta_2-1},
\end{align}
\begin{align}
f_2(w_t,s_t |\textbf{w},\textbf{s})=&\dfrac{(n-n_r)!}{(n-n_r-t+r)!}\sum\limits_{j_2=0}^{t-r-1}c_{j_2}(t-r-1)\big(1+w_{t}^{\theta_4}\big)^{-\theta_3(n-n_r-t+r+j_2+1)-1}\nonumber\\
&\big(1+w_{r}^{\theta_4}\big)^{-\theta_3(n_r-n+t-r-j_2-1)}\theta_3\theta_4 w_{t}^{\theta_4-1},
\end{align}
and also, using the binomial expansion for the terms $\big[(1+w_{r}^{\theta_2})^{-\theta_1}-(1+w_{t}^{\theta_2})^{-\theta_1}\big]^{m_{t-1}-m_r}$ and $\big[(1+w_{r}^{\theta_4})^{-\theta_3}-(1+w_{t}^{\theta_4})^{-\theta_3}\big]^{n_{t-1}-n_r}$, \autoref{CDF3:W_{r+j}} can be rewritten in the following form

\begin{align}
f_3(w_t,s_t |\textbf{w},\textbf{s})=&(m-m_r)!(n-n_r)!
\sum_{s_{r+1}=0}^{1} \cdots \sum_{s_{t}=0}^{1}\sum_{j_3=0}^{m_{t-1}-m_r}\sum_{j_4=0}^{n_{t-1}-n_r}\dfrac{c_{j_3}(m_{t-1}-m_r)c_{j_4}(n_{t-1}-n_r)}{(m-m_{t})!(n-n_{t})!}\nonumber\\
&\times\big[1+w_{t}^{\theta_2}\big]^{-\theta_1(m-m_{t-1}+j_3)-s_t}(1+w_{r}^{\theta_2})^{-\theta_1(m_{t-1}-j_3-m)}(1+w_{r}^{\theta_4})^{-\theta_3(n_{t-1}-j_4-n)}\nonumber\\
&\times\big[1+w_{t}^{\theta_4}\big]^{-\theta_3(n-n_{t-1}+j_4)-(1-s_t)}\theta_1^{s_t}\theta_2^{s_t}\theta_3^{1-s_t}\theta_4^{1-s_t} w_{t}^{-s_t(\theta_4-\theta_2)+\theta_4-1},
\end{align}
where $c_j(i)=\dfrac{(-1)^{j}}{j!(i-j)!}$, for $j=0,1,2,...,i$.

In the following subsections, BUP and Bayesian Prediction methods will be discussed.

\subsection{Best unbiased predictor}
This subsection discusses the application of the BUP in predicting future failures $W_{t}$. The BUP, denoted as $\widehat{W}_{t}$, is considered the optimal predictor for $W_{t}$ if the prediction error $(\widehat{W}_{t}-W_{t})$ has a mean of zero, and its variance, \text{var}$(\widehat{W}_{t}-W_{t})$, is the smallest or equal to all other unbiased estimators. Therefore, the BUP of $W_t$ using \autoref{cpdf:w_{r+j}} is given by

\begin{eqnarray}
\label{prediction:w_{T}}
\widehat{W}_{t}=E\big(W_{t},S_{t}|\textbf{w},\textbf{s}\big)=\int_{w_r}^{\infty}w_{t}f(w_{t},s_{t} |\textbf{w},\textbf{s})dw_{t}=
\begin{cases}
\widehat{W}_{t.1},\quad{m_r} < m,\quad{n_r} = n,\\
\widehat{W}_{t.2},\quad{m_r} = m,\quad{n_r} < n,\\
\widehat{W}_{t.3},\quad{m_r} < m,\quad{n_r} < n,
\end{cases}
\end{eqnarray}
where 
\begin{equation}\label{E1:W_{r+j}}
\widehat{W}_{t.1}=\dfrac{1}{Beta(m-m_r-t+r+1,t-r)}\int_{0}^{1}\big[u_{1}^{-1/\theta_1}(1+w_{r}^{\theta_2})-1\big]^{1/\theta_2}u_{1}^{m-m_r-t+r}(1-u_1)^{t-r-1}du_1,
\end{equation}
and $u_1=\Big(\dfrac{1+w_{t}^{\theta_2}}{1+w_{r}^{\theta_2}}\Big)^{-\theta_1}|(\textbf{w}, \textbf{s})$ is $Beta(m-m_r-t+r+1,t-r)$ distribution. Also,
\begin{equation}\label{E2:W_{r+j}}
\widehat{W}_{t.2}=\dfrac{1}{ Beta(n-n_r-t+r+1,t-r)}\int_{0}^{1}\big[u_{2}^{-1/\theta_3}(1+w_{r}^{\theta_4})-1\big]^{1/\theta_4}u_{2}^{n-n_r-t+r}(1-u_2)^{t-r-1}du_2,
\end{equation}
and $u_2=\Big(\dfrac{1+w_{t}^{\theta_4}}{1+w_{r}^{\theta_4}}\Big)^{-\theta_3}|(\textbf{w}, \textbf{s})$ is $Beta(n-n_r-t+r+1,t-r)$ distribution. Finally, the third part of the \autoref{prediction:w_{T}} will be the following form 
\begin{align}\label{E3:W_{r+j}}
 \widehat{W}_{t.3}=&\int_{w_r}^{\infty}w_{t}f_3(w_{t},s_{t} |\textbf{w}, \textbf{s})dw_{t}=\sum_{s_{r+1}=0}^{1} \cdots \sum_{s_{t-1}=0}^{1}\nonumber\\
 &\dfrac{(m-m_r)!(n-n_r)!\theta_3\theta_4(1+w_{r}^{\theta_2})^{-\theta_1(m_r-m)}(1+w_{r}^{\theta_4})^{-\theta_3(n_r-n)}}{(m_{t-1}-m_r)!(m-m_{t-1})!(n_{t-1}-n_r)!(n-n_{t-1}-1)!}\nonumber\\
 &\times\Bigg(\int_{w_r}^{\infty}w_{t}^{\theta_4}(1+w_{t}^{\theta_2})^{-\theta_1(m-m_{t-1})}(1+w_{t}^{\theta_4})^{-\theta_3(n-n_{t-1})-1}\nonumber\\
 &\times\big[(1+w_{r}^{\theta_2})^{-\theta_1}-(1+w_{t}^{\theta_2})^{-\theta_1}\big]^{m_{t-1}-m_r}\big[(1+w_{r}^{\theta_4})^{-\theta_3}-(1+w_{t}^{\theta_4})^{-\theta_3}\big]^{n_{t-1}-n_r}dw_t\Bigg)\nonumber\\
 &+\sum_{s_{r+1}=0}^{1} \cdots \sum_{s_{t-1}=0}^{1}\dfrac{(m-m_r)!(n-n_r)!\theta_1\theta_2(1+w_{r}^{\theta_2})^{-\theta_1(m_r-m)}(1+w_{r}^{\theta_4})^{-\theta_3(n_r-n)}}{(m_{t-1}-m_r)!(m-m_{t-1}-1)!(n_{t-1}-n_r)!(n-n_{t-1})!}\nonumber\\
  &\times\Bigg(\int_{w_r}^{\infty}w_{t}^{\theta_2}(1+w_{t}^{\theta_2})^{-\theta_1(m-m_{t-1})-1}(1+w_{t}^{\theta_4})^{-\theta_3(n-n_{t-1})}\nonumber\\
&\times \big[(1+w_{r}^{\theta_2})^{-\theta_1}-(1+w_{t}^{\theta_2})^{-\theta_1}\big]^{m_{t-1}-m_r}\big[(1+w_{r}^{\theta_4})^{-\theta_3}-(1+w_{t}^{\theta_4})^{-\theta_3}\big]^{n_{t-1}-n_r}dw_t\Bigg).
\end{align}

To compute the BUP using \autorefs{E1:W_{r+j}}, \ref{E2:W_{r+j}}, and \ref{E3:W_{r+j}}, the unknown parameters $\theta_1$, $\theta_2$, $\theta_3$ and $\theta_4$ are substituted with their respective MLEs obtained from the EM method. 

\subsection{Bayesian Prediction}
In this subsection, Bayesian prediction of future failure $W_{t}$ is discussed. For this purpose, first the Bayesian prediction function of $W_{t}$ given $(\textbf{w},\textbf{s})$ is computed in the following form
\begin{eqnarray}
\label{bayse prediction:w_{r+j}}
f^{*}(w_{t},s_{t}|\textbf{w,s})= 
\begin{cases}
f_{1}^{*}(w_{t},s_{t}|\textbf{w},\textbf{s}),\quad{m_r} < m,\quad{n_r} = n,\\
f_{2}^{*}(w_{t},s_{t}|\textbf{w},\textbf{s}),\quad{m_r} = m,\quad{n_r} < n,\\
f_{3}^{*}(w_{t},s_{t}|\textbf{w},\textbf{s}),\quad{m_r} < m,\quad{n_r} < n,
 \end{cases}
\end{eqnarray}
where
\begin{equation}\label{BPD1}
f_{1}^{*}(w_{t},s_{t}|\textbf{w},\textbf{s})=\int_{0}^{\infty}\int_{0}^{\infty}\int_{0}^{\infty}\int_{0}^{\infty}f_{1}(w_{t},s_{t}|\textbf{w},\textbf{s})\pi(\boldsymbol{\theta}|\textbf{w},\textbf{s})d\theta_1 d\theta_2 d\theta_3 d\theta_4,
\end{equation}
\begin{equation}\label{BPD2}
f_{2}^{*}(w_{t},s_{t}|\textbf{w},\textbf{s})=\int_{0}^{\infty}\int_{0}^{\infty}\int_{0}^{\infty}\int_{0}^{\infty}f_{2}(w_{t},s_{t}|\textbf{w},\textbf{s})\pi(\boldsymbol{\theta}|\textbf{w},\textbf{s})d\theta_1 d\theta_2 d\theta_3 d\theta_4,
\end{equation}
and
\begin{equation}\label{BPD3}
f_{3}^{*}(w_{t},s_{t}|\textbf{w},\textbf{s})=\int_{0}^{\infty}\int_{0}^{\infty}\int_{0}^{\infty}\int_{0}^{\infty}f_{3}(w_{t},s_{t}|\textbf{w},\textbf{s})\pi(\boldsymbol{\theta}|\textbf{w},\textbf{s})d\theta_1 d\theta_2 d\theta_3 d\theta_4,
\end{equation}
so $f_{1}(w_{t}|\textbf{w},\textbf{s})$, $f_{2}(w_{t}|\textbf{w},\textbf{s})$, $f_{3}(w_{t}|\textbf{w},\textbf{s})$ and $\pi(\boldsymbol{\theta}|\textbf{w},\textbf{s})$ are presented in \autorefs{CDF1:W_{r+j}}-\ref{CDF3:W_{r+j}} and \ref{importance:informative}, respectively. Now, the Bayesian predictive estimate of $W_{t}$ under the SE loss function is given by
\begin{eqnarray}\label{SE.PREDICT}
\widehat{W}_{t.SE}=E\big(W_t,S_t|\textbf{w},\textbf{s}\big)= 
\begin{cases}
\widehat{W}_{t.SE.1},\quad{m_r} < m,\quad{n_r} = n,\\
\widehat{W}_{t.SE.2},\quad{m_r} = m,\quad{n_r} < n,\\
\widehat{W}_{t.SE.3},\quad{m_r} < m,\quad{n_r} < n,
 \end{cases}
\end{eqnarray}
where
\begin{align}\label{E1:sel}
\widehat{W}_{t.SE.1}&=\int_{w_r}^{\infty}w_{t}f_{1}^{*}(w_{t},s_{t}|\textbf{w},\textbf{s})dw_{t}=\int_{0}^{\infty}\int_{0}^{\infty}\int_{0}^{\infty}\int_{0}^{\infty} I_{1.SE}(\theta_1,\theta_2|\textbf{w},\textbf{s})\pi_1(\boldsymbol{\theta}|\textbf{w},\textbf{s})d\theta_1 d\theta_2 d\theta_3d\theta_4\nonumber\\
&=E\big[I_{1.SE}(\theta_1,\theta_2|\textbf{w},\textbf{s})\big],
\end{align}
so that $I_{1.SE}(\theta_1,\theta_2|\textbf{w},\textbf{s})=\int_{w_r}^{\infty}w_{t}f_{1}(w_{t},s_{t}|\textbf{w,s})dw_{t}=\widehat{W}_{t.1}$, which is calculated in \autoref{E1:W_{r+j}}. Similarly, $\widehat{W}_{t.SE.2}$ can be calculated in the following form

\begin{align}\label{E2:sel}
\widehat{W}_{t.SE.2}&=\int_{w_r}^{\infty}w_{t}f_{2}^{*}(w_{t},s_{t}|\textbf{w},\textbf{s})dw_{t}=\int_{0}^{\infty}\int_{0}^{\infty}\int_{0}^{\infty}\int_{0}^{\infty} I_{2.SE}(\theta_3,\theta_4|\textbf{w},\textbf{s})\pi(\boldsymbol{\theta}|\textbf{w},\textbf{s})d\theta_1 d\theta_2 d\theta_3 d\theta_4\nonumber\\
&=E\big[I_{2.SE}(\theta_3,\theta_4|\textbf{w},\textbf{s})\big],
\end{align}
and $I_{2.SE}(\theta_3,\theta_4|\textbf{w},\textbf{s})=\int_{w_r}^{\infty}w_{t}f_{2}(w_{t},s_{t}|\textbf{w,s})dw_{t}=\widehat{W}_{t.2}$, that is computed in \autoref{E2:W_{r+j}}. Finally, $\widehat{W}_{t.SE.3}$ will be computed in the following form
\begin{align}\label{E3:sel}
\widehat{W}_{t.SE.3}&=\int_{w_r}^{\infty}w_{t}f_{3}^{*}(w_{t},s_{t}|\textbf{w},\textbf{s})dw_{t}=\int_{0}^{\infty}\int_{0}^{\infty}\int_{0}^{\infty}\int_{0}^{\infty} I_{3.SE}(\boldsymbol{\theta}|\textbf{w},\textbf{s})\pi(\boldsymbol{\theta}|\textbf{w},\textbf{s})d\theta_1 d\theta_2 d\theta_3 d\theta_4\nonumber\\
&=E\big[I_{3.SE}(\boldsymbol{\theta}|\textbf{w},\textbf{s})\big],
\end{align}
and $I_{3.SE}(\boldsymbol{\theta}|\textbf{w},\textbf{s})=\int_{w_r}^{\infty}w_{t}f_{3}^{*}(w_{t},s_{t}|\textbf{w},\textbf{s})dw_{t}=\widehat{W}_{t.3}$ that is given in \autoref{E3:W_{r+j}}. It is evident that \autorefs{E1:sel}-\ref{E3:sel} do not have a closed-form solution. Consequently, the numerical approximation is required to calculate $\widehat{W}_{t.SE}$.  For this purpose, we utilize the importance sampling technique discussed in \autoref{sec2}.  We assume that the $\theta_{1j},\theta_{2j},\theta_{3j},\theta_{4j}$ ($j=1,2,...,D$) are generated from $\pi(\boldsymbol{\theta}|\textbf{w},\textbf{s})$, presented in \autoref{importance:informative}. Therefore, based on \autoref{SE.PREDICT}, the following form is obtained as
\begin{eqnarray}\label{zhat:sel}
\widehat{W}_{t.SE}=
\begin{cases}
\dfrac{\sum\limits_{j=1}^{D}I_{1.SE}(\theta_{1j},\theta_{2j}|\textbf{w},\textbf{s})\Delta(\theta_{2j},\theta_{4j})}{\sum\limits_{j=1}^{D}\Delta(\theta_{2j},\theta_{4j})},\quad m_r<m,\quad n_r=n,\\
\dfrac{\sum\limits_{j=1}^{D}I_{2.SE}(\theta_{3j},\theta_{4j}|\textbf{w},\textbf{s})\Delta(\theta_{2j},\theta_{4j})}{\sum\limits_{j=1}^{D}\Delta(\theta_{2j},\theta_{4j})},\quad m_r=m,\quad n_r<n,\\
\dfrac{\sum\limits_{j=1}^{D}I_{3.SE}(\theta_{1j},\theta_{2j},\theta_{3j},\theta_{4j}|\textbf{w},\textbf{s})\Delta(\theta_{2j},\theta_{4j})}{\sum\limits_{j=1}^{D}\Delta(\theta_{2j},\theta_{4j})},\quad m_r<m,\quad n_r<n,
\end{cases}
\end{eqnarray}
so $\Delta(\theta_{2j},\theta_{4j})$ is define in \autoref{Delta}. Bayesian predictive estimate of $W_{t}$ under LINEX loss function is obtained as 
\begin{eqnarray}\label{LINEX:Predictive}
\widehat{W}_{t.LINEX}=-\dfrac{1}{v}\ln E[e^{-vw_t}|\textbf{w},\textbf{s}]
=\begin{cases}
\widehat{W}_{t.LINEX.1}, \quad m_r<m, \quad n_r=n,\\
\widehat{W}_{t.LINEX.2},\quad m_r=m,\quad n_r<n,\\
\widehat{W}_{t.LINEX.3},\quad m_r<m,\quad n_r<n,
\end{cases}
\end{eqnarray}
where
\begin{align}\label{E1:linex}
\widehat{W}_{t.LINEX.1}=&-\dfrac{1}{v}\ln\int_{w_r}^{\infty}e^{-v w_{t}}f_{1}^{*}(w_{t},s_{t}|\textbf{w,s})dw_{t}=-\dfrac{1}{v}\ln\int_{0}^{\infty}\int_{0}^{\infty}\int_{0}^{\infty}\int_{0}^{\infty} I_{1.LINEX}(\theta_1,\theta_2|\textbf{w},\textbf{s})\nonumber\\
&\times\pi(\boldsymbol{\theta}|\textbf{w},\textbf{s})d\theta_1 d\theta_2 d\theta_3 d\theta_4=-\dfrac{1}{v}\ln E(I_{1.LINEX}(\theta_1,\theta_2)|\textbf{w,s}),
\end{align}
\begin{align*}
I_{1.LINEX}(\theta_1,\theta_2|\textbf{w},\textbf{s})=&\int_{w_r}^{\infty}e^{-vw_{t}}f_{1}(w_{t},s_{t}|\textbf{w}, \textbf{s})dw_{t}=\dfrac{1}{Beta(m-m_r-t+r+1,t-r)}\nonumber\\
&\times\int_{0}^{1}e^{-v(u_{1}^{-1/\theta_1}(1+w_{r}^{\theta_2})-1)^{1/\theta_2}}u_{1}^{m-m_r-t+r}(1-u_1)^{t-r-1}du_1,
\end{align*}
and $u_1=\Big[\dfrac{1+w_{t}^{\theta_2}}{1+w_{r}^{\theta_2}}\Big]^{-\theta_1}$. Also
\begin{align}\label{E2:linex}
\widehat{W}_{t.LINEX.2}=&-\dfrac{1}{v}\ln \int_{w_r}^{\infty}e^{-v w_{t}}f_{2}^{*}(w_{t},s_{t}|\textbf{w},\textbf{s})dw_{t}=-\dfrac{1}{v}\ln \int_{0}^{\infty}\int_{0}^{\infty}\int_{0}^{\infty}\int_{0}^{\infty} I_{2.LINEX}(\theta_3,\theta_4|\textbf{w},\textbf{s})\nonumber\\
&\times\pi(\boldsymbol{\theta}|\textbf{w},\textbf{s})d\theta_1 d\theta_2 d\theta_3 d\theta_4=-\dfrac{1}{v}\ln E(I_{2.LINEX}(\theta_3,\theta_4)|\textbf{w,s}),
\end{align}
\begin{align*}
 I_{2.LINEX}(\theta_3,\theta_4|\textbf{w},\textbf{s})=&\int_{w_r}^{\infty}e^{-v w_{t}}f_{2}(w_{t},s_{t}|\textbf{w},\textbf{s})dw_{t}=\dfrac{1}{ Beta(n-n_r-t+r+1,t-r)}\nonumber\\
&\times\int_{0}^{1}e^{-v(u_{2}^{-1/\theta_3}(1+w_{r}^{\theta_4})-1)^{1/\theta_4}}u_{2}^{n-n_r-t+r}(1-u_2)^{t-r-1}du_2,
\end{align*}
and $u_2=\Big[\dfrac{1+w_{t}^{\theta_4}}{1+w_{r}^{\theta_4}}\Big]^{-\theta_3}$. Finally, the third part of the \autoref{LINEX:Predictive} will be computed as the following form
\begin{align}\label{E3:linex}
\widehat{W}_{t.LINEX.3}=&-\dfrac{1}{v}\ln\int_{w_r}^{\infty}e^{-v w_{t}}f_{3}^{*}(w_{t},s_{t}|\textbf{w,s})dw_{t}=-\dfrac{1}{v}\ln\int_{0}^{\infty}\int_{0}^{\infty}\int_{0}^{\infty}\int_{0}^{\infty} I_{3.LINEX}(\boldsymbol{\theta}|\textbf{w},\textbf{s})\nonumber\\
&\times\pi(\boldsymbol{\theta}|\textbf{w},\textbf{s})d\theta_1 d\theta_2 d\theta_3 d\theta_4=-\dfrac{1}{v}\ln E(I_{3.LINEX}(\boldsymbol{\theta})|\textbf{w,s}),
\end{align}
and $I_{3.LINEX}(\theta|\textbf{w},\textbf{s})=\int_{w_r}^{\infty}e^{-vw_{t}}f_3(w_{t},s_{t}|\textbf{w,s})dw_{t}$. Based on \autoref{LINEX:Predictive} and using importance sampling technique, the predictive value for $W_{t}$ under LINEX loss function is given by 
\begin{eqnarray}\label{zhat:linex}
\widehat{W}_{t.LINEX}=
\begin{cases}
-\dfrac{1}{v}\ln\Bigg(\dfrac{\sum\limits_{j=1}^{D}e^{-vI_{1.LINEX}(\theta_{1j},\theta_{2j}|\textbf{w},\textbf{s})}\Delta(\theta_{2j},\theta_{4j})}{\sum\limits_{j=1}^{D}\Delta(\theta_{2j},\theta_{4j})}\Bigg),\quad m_r<m,\quad n_r=n,\\
-\dfrac{1}{v}\ln\Bigg(\dfrac{\sum\limits_{j=1}^{D}e^{-vI_{2.LINEX}(\theta_{3j},\theta_{4j}|\textbf{w},\textbf{s})}\Delta(\theta_{2j},\theta_{4j})}{\sum\limits_{j=1}^{D}\Delta(\theta_{2j},\theta_{4j})}\Bigg),\quad m_r=m,\quad n_r<n,\\
-\dfrac{1}{v}\ln\Bigg(\dfrac{\sum\limits_{j=1}^{D}e^{-vI_{3.LINEX}(\theta_{1j},\theta_{2j},\theta_{3j},\theta_{4j}|\textbf{w},\textbf{s})}\Delta(\theta_{2j},\theta_{4j})}{\sum\limits_{j=1}^{D}\Delta(\theta_{2j},\theta_{4j})}\Bigg),\quad m_r<m,\quad n_r<n,\\.
\end{cases}
\end{eqnarray}

Subsequently, $W_{t}$ under the GE loss function can be predicted as
 \begin{eqnarray}\label{GE:Predictive}
\widehat{W}_{t.GE}=\big[E(w_{t}^{-k}|w,s)\big]^{-\frac{1}{k}}
=\begin{cases}
\widehat{W}_{t.GE.1},\quad m_r<m, \quad n_r=n,\\
\widehat{W}_{t.GE.2},\quad  m_r=m,\quad  n_r<n,\\
\widehat{W}_{t.GE.3}, \quad m_r<m, \quad n_r<n,
\end{cases}
\end{eqnarray}
where
\begin{align}\label{E1:gelf}
\widehat{W}_{t.GE.1}=&\Bigg[\int_{w_r}^{\infty}w_{t}^{-k}f_{1}^{*}(w_{t},s_{t}|\textbf{w,s})dw_{t}\Bigg]^{-\frac{1}{k}}=\Bigg[\int_{0}^{\infty}\int_{0}^{\infty}\int_{0}^{\infty}\int_{0}^{\infty}I_{1.GE}(\theta_1,\theta_2|\textbf{w},\textbf{s})\nonumber\\
&\times\pi(\boldsymbol{\theta}|\textbf{w},\textbf{s})d\theta_1d\theta_2d\theta_3d\theta_4\Bigg]^{-\frac{1}{k}}=\big[E(I_{1.GE}(\theta_1,\theta_2|\textbf{w},\textbf{s})\big]^{-\frac{1}{k}},
\end{align}
\begin{align*}
I_{1.GE}(\theta_1,\theta_2|\textbf{w},\textbf{s})=&\int_{w_r}^{\infty}w_{t}^{-k}f_{1}(w_{t},s_{t}|\textbf{w}, \textbf{s})dw_{t}=\dfrac{1}{Beta(m-m_r-t+r+1,t-r)}\nonumber\\
&\times\int_{0}^{1}(u_{1}^{1/\theta_1}(1+w_{r}^{\theta_2})-1)^{-k/\theta_2}u_{1}^{m-m_r-t+r}(1-u_1)^{t-r-1}du_1,
\end{align*}
and $u_1=\Big[\dfrac{1+w_{t}^{\theta_2}}{1+w_{r}^{\theta_2}}\Big]^{-\theta_1}$. Also
\begin{align}\label{E2:gelf}
\widehat{W}_{t.GE.2}=&\Bigg[\int_{w_r}^{\infty}w_{t}^{-k}f_{2}^{*}(w_{t},s_{t}|\textbf{w,s})dw_{t}\Bigg]^{-\frac{1}{k}}=\Bigg[\int_{0}^{\infty}\int_{0}^{\infty}\int_{0}^{\infty}\int_{0}^{\infty}I_{2.GE}(\theta_3,\theta_4|\textbf{w},\textbf{s})\nonumber\\
&\times\pi(\boldsymbol{\theta}|\textbf{w},\textbf{s})d\theta_1d\theta_2d\theta_3d\theta_4\Bigg]^{-\frac{1}{k}}=\big[E(I_{2.GE}(\theta_3,\theta_4|\textbf{w},\textbf{s})\big]^{-\frac{1}{k}},
\end{align}
\begin{align*}
I_{2.GE}(\theta_3,\theta_4)=&\int_{w_r}^{\infty} w_{t}^{-k}f_{2}(w_{t},s_{t}|\textbf{w},\textbf{s})dw_{t}=\dfrac{1}{ Beta(n-n_r-t+r+1,t-r)}\nonumber\\
&\times\int_{0}^{1}(u_{2}^{-1/\theta_3}(1+w_{r}^{\theta_4})-1)^{-k/\theta_4}u_{2}^{n-n_r-t+r}(1-u_2)^{t-r-1}du_2,
\end{align*}
and $u_2=\Big[\dfrac{1+w_{t}^{\theta_4}}{1+w_{r}^{\theta_4}}\Big]^{-\theta_3}$. Finally, we have 

\begin{align}\label{E3:gelf}
\widehat{W}_{t.GE.3}=&\Bigg[\int_{w_r}^{\infty}w_{t}^{-k}f_{3}^{*}(w_{t},s_{t}|\textbf{w,s})dw_{t}\Bigg]^{-\frac{1}{k}}=\Bigg[\int_{0}^{\infty}\int_{0}^{\infty}\int_{0}^{\infty}\int_{0}^{\infty}I_{3.GE}(\boldsymbol{\theta}|\textbf{w},\textbf{s})\nonumber\\
&\times\pi(\boldsymbol{\theta}|\textbf{w},\textbf{s})d\theta_1d\theta_2d\theta_3d\theta_4\Bigg]^{-\frac{1}{k}}=\big[E(I_{3.GE}(\boldsymbol{\theta}|\textbf{w},\textbf{s})\big]^{-\frac{1}{k}},
\end{align}
and $I_{3.GE}(\boldsymbol{\theta})=\int_{w_r}^{\infty}w_{t}^{-k}f_3(w_{t},s_{t}|\textbf{w},\textbf{s})dw_{t}$. Using importance sampling method and \autoref{GE:Predictive}, Bayesian predictive estimate of $W_{t}$ under GE loss function is given by 
\begin{eqnarray}\label{zhat:gelf}
\hat{W}_{t.GE}=
\begin{cases}
\Bigg(\dfrac{\sum\limits_{j=1}^{D}\big[I_{1.GE}(\theta_{1j},\theta_{2j}|\textbf{w},\textbf{s})\big]^{-k}\Delta(\theta_{2j},\theta_{4j})}{\sum\limits_{j=1}^{D}\Delta(\theta_{2j},\theta_{4j})}\Bigg)^{-\frac{1}{k}},\quad m_r<m,\quad n_r=n,\\
\Bigg(\dfrac{\sum\limits_{j=1}^{D}\big[I_{2.GE}(\theta_{3j},\theta_{4j}|\textbf{w},\textbf{s})\big]^{-k}\Delta(\theta_{2j},\theta_{3j})}{\sum\limits_{j=1}^{D}\Delta(\theta_{2j},\theta_{4j})}\Bigg)^{-\frac{1}{k}},\quad m_r=m,\quad n_r<n,\\
\Bigg(\dfrac{\sum\limits_{j=1}^{D}\big[I_{3.GE}(\theta_{1j},\theta_{2j},\theta_{3j},\theta_{4j}|\textbf{w},\textbf{s})\big]^{-k}\Delta(\theta_{2j},\theta_{4j})}{\sum\limits_{j=1}^{D}\Delta(\theta_{2j},\theta_{4j})}\Bigg)^{-\frac{1}{k}},\quad m_r<m,\quad n_r<n.\\
\end{cases}
\end{eqnarray}

To calculate the Bayesian prediction for $W_t$ using \autorefs{zhat:sel}, \ref{zhat:linex}, and \ref{zhat:gelf}, the vector of unknown parameters $\boldsymbol{\theta}$ is substituted with its corresponding Bayesian estimates obtained from the importance sampling method. 

\subsection{Prediction Intervals}
This subsection explores the estimation of prediction intervals for future failures $W_{t}$ using classical and Bayesian methods, which are described in following subsections.

\subsubsection{Classical Prediction Intervals }\label{subsec3}
Using the conditional density function given by \autoref{cpdf:w_{r+j}}, it is possible to derive the prediction survival function for given $\xi$ in the following form

\begin{eqnarray}\label{p.s.f}
\bar{F}(w_t>\xi|\textbf{w},\textbf{s})=
\begin{cases}
\bar{F_1}(w_t>\xi|\textbf{w},\textbf{s}),\quad m_r<m,\quad n_r=n,\\
\bar{F_2}(w_t>\xi|\textbf{w},\textbf{s}),\quad m_r=m,\quad n_r<n,\\
\bar{F_3}(w_t>\xi|\textbf{w},\textbf{s}),\quad m_r<m,\quad n_r<n,
\end{cases}
\end{eqnarray}
where 
\begin{align}\label{p.s.f.1}
\bar{F_1}(w_t>\xi|\textbf{w},\textbf{s})=&\int_{\xi}^{\infty} f_1(w_{t},s_{t}|\textbf{w},\textbf{s})dw_t=\dfrac{(m-m_r)!}{(m-m_r-t+r)!}\nonumber\\
&\times\sum_{j_1=0}^{t-r-1}\dfrac{c_{j_1}(t-r-1)}{m-m_r-t+r+j_1+1}\Big(\dfrac{1+\xi^{\theta_2}}{1+w_r^{\theta_2}}\Big)^{-\theta_1(m-m_r-t+r+j_1+1)},
\end{align}
\begin{align}\label{p.s.f.2}
\bar{F_2}(w_t>\xi|\textbf{w},\textbf{s})=&\int_{\xi}^{\infty} f_2(w_{t},s_{t}|\textbf{w},\textbf{s})dw_t=\dfrac{(n-n_r)!}{(n-n_r-t+r)!}\nonumber\\
&\times\sum_{j_2=0}^{t-r-1}\dfrac{c_{j_2}(t-r-1)}{n-n_r-t+r+j_2+1}\Big(\dfrac{1+\xi^{\theta_4}}{1+w_{r}^{\theta_4}}\Big)^{-\theta_3(n-n_r-t+r+j_2+1)},
\end{align}
and
\begin{align}\label{p.s.f.3}
\bar{F_3}(w_t>\xi|\textbf{w},\textbf{s})=&\int_{\xi}^{\infty}f_3(w_{t},s_{t}|\textbf{w},\textbf{s})dw_t=(m-m_r)!(n-n_r)!\nonumber\\
&\times\sum_{s_{r+1}=0}^{1} \cdots \sum_{s_{t-1}=0}^{1}\sum_{j_3=0}^{m_{t-1}-m_r}\sum_{j_4=0}^{n_{t-1}-n_r}(1+w_{r}^{\theta_2})^{-\theta_1(m_{t-1}-m-j_3)}(1+w_{r}^{\theta_4})^{-\theta_3(n_{t-1}-n-j_4)}\nonumber\\
&\times\Bigg\{\dfrac{c_{j_3}(m_{t-1}-m_r)c_{j_4}(n_{t-1}-n_r)}{(m-m_{t-1})!(n-n_{t-1}-1)!}\int_{\xi}^{\infty}\Big(\theta_3\theta_4(1+w_t^{\theta_2})^{-\theta_1(m-m_{t-1}+j_3)}\nonumber\\
&\times(1+w_t^{\theta_4})^{-\theta_3(n-n_{t-1}+j_4)-1}w_t^{\theta_4-1}\Big)dw_t+\dfrac{c_{j_3}(m_{t-1}-m_r)c_{j_4}(n_{t-1}-n_r)}{(m-m_{t-1}-1)!(n-n_{t-1})!} \nonumber\\
&\times\int_{\xi}^{\infty}\theta_1\theta_2(1+w_t^{\theta_2})^{-\theta_1(m-m_{t-1}+j_3)-1}
(1+w_t^{\theta_4})^{-\theta_3(n-n_{t-1}+j_4)}w_t^{\theta_2-1}dw_t.\Bigg\} 
\end{align}

There is no closed form for \autoref{p.s.f.3}. The $100(1 - \alpha )\% $ equal tail classical prediction intervals for  $W_{t}$ are calculated in the following form
\begin{eqnarray}\label{L.C.P.I }
\bar{F}(w_t>L_{w_{t}}|\textbf{w,s})=
\begin{cases}
\bar{F_1}(w_t>L_{W_{t}}|\textbf{w},\textbf{s})=1-\frac{\alpha}{2},\quad m_r<m,\quad n_r=n,\\
\bar{F_2}(w_t>L_{W_{t}}|\textbf{w},\textbf{s})=1-\frac{\alpha}{2},\quad m_r=m,\quad n_r<n,\\
\bar{F_3}(w_t>L_{W_{t}}|\textbf{w},\textbf{s})=1-\frac{\alpha}{2},\quad m_r<m,\quad n_r<n,
\end{cases}
\end{eqnarray}
 and
 \begin{eqnarray}\label{U.C.P.I}
\bar{F}(w_t>U_{W_{t}}|\textbf{w,s})=
\begin{cases}
\bar{F_1}(w_t>U_{W_{t}}|\textbf{w},\textbf{s})=\frac{\alpha}{2},\quad m_r<m,\quad n_r=n,\\
\bar{F_2}(w_t>U_{W_{t}}|\textbf{w},\textbf{s})=\frac{\alpha}{2},\quad m_r=m,\quad n_r<n,\\
\bar{F_3}(w_t>U_{W_{t}}|\textbf{w},\textbf{s})=\frac{\alpha}{2},\quad m_r<m,\quad n_r<n,
\end{cases}
\end{eqnarray}
as $L_{w_{t}}$ and $U_{w_{t}}$ represent the lower and upper limits for $W_{t}$, respectively. By numerically solving \autorefs{L.C.P.I } and \ref{U.C.P.I}, we can obtain lower and upper bounds for $W_t$, It should be noted that in these equations, $\theta_1$, $\theta_2$, $\theta_3$, and $\theta_4$ are replaced with their corresponding MLEs, respectively.

\subsubsection{Bayesian Prediction Intervals}\label{subsec4}
Based on \autoref{bayse prediction:w_{r+j}}, the Bayesian posterior survival function for given $\xi$ can be written in the following form
\begin{eqnarray}\label{B.P.S.F}
\bar{F}^*(w_t>\xi|\textbf{w,s})=
\begin{cases}
\bar{F}_{1}^*(w_t>\xi|\textbf{w},\textbf{s}),\quad m_r<m,\quad n_r=n,\\
\bar{F}_{2}^*(w_t>\xi|\textbf{w},\textbf{s}),\quad m_r=m,\quad n_r<n,\\
\bar{F}_{3}^*(w_t>\xi|\textbf{w},\textbf{s}),\quad m_r<m,\quad n_r<n,
\end{cases}
\end{eqnarray}
where
\begin{align}\label{B.P.S.F.1}
\bar{F}_{1}^*(w_t>\xi|\textbf{w},\textbf{s})=&\int_{\xi}^{\infty} f_{1}^{*}(w_{t},s_{t}|\textbf{w},\textbf{s})dw_t=\dfrac{(m-m_r)!}{(m-m_r-t+r)!}\sum_{j_1=0}^{t-r-1}\dfrac{c_{j_1}(t-r-1)}{m-m_r-t+r+j_1+1}\nonumber\\
&\times\int_{0}^{\infty}\int_{0}^{\infty}\int_{0}^{\infty}\int_{0}^{\infty}\Big[\dfrac{1+\xi^{\theta_2}}{1+w_{r}^{\theta_2}}\Big]^{-\theta_1(m-m_r-t+r+j_1+1)}
\pi(\boldsymbol{\theta}|\textbf{w}, \textbf{s})d\theta_1d\theta_2d\theta_3d\theta_4,
\end{align}
\begin{align}\label{B.P.S.F.2}
\bar{F}_{2}^*(w_t>\xi|\textbf{w},\textbf{s})=&\int_{\xi}^{\infty} f_{2}^{*}(w_{t},s_{t}|\textbf{w},\textbf{s})dw_t=\dfrac{(n-n_r)!}{(n-n_r-t+r)!}\sum_{j_2=0}^{t-r-1}\dfrac{c_{j_2}(t-r-1)}{n-n_r-t+r+j_2+1}\nonumber\\
&\times\int_{0}^{\infty}\int_{0}^{\infty}\int_{0}^{\infty}\int_{0}^{\infty}\Big[\dfrac{1+\xi^{\theta_4}}{1+w_{r}^{\theta_4}}\Big]^{-\theta_3(n-n_r-t+r+j_2+1)}
\pi(\boldsymbol{\theta}|\textbf{w}, \textbf{s})d\theta_1d\theta_2d\theta_3d\theta_4,
\end{align}
\begin{align}\label{B.P.S.F.3}
\bar{F_3}(w_t>\xi|\textbf{w},\textbf{s})=&\int_{\xi}^{\infty}f_3(w_{t},s_{t}|\textbf{w},\textbf{s})dw_t=(m-m_r)!(n-n_r)!\int_{0}^{\infty}\int_{0}^{\infty}\int_{0}^{\infty}\int_{0}^{\infty}\Bigg(\nonumber\\
&\sum_{s_{r+1}=0}^{1}\ldots \sum_{s_{t-1}=0}^{1}\sum_{j_3=0}^{m_{t-1}-m_r}\sum_{j_4=0}^{n_{t-1}-n_r}(1+w_{r}^{\theta_2})^{-\theta_1(m_{t-1}-m-j_3)}(1+w_{r}^{\theta_4})^{-\theta_3(n_{t-1}-n-j_4)}\nonumber\\
&\times\Bigg\{\dfrac{c_{j_3}(m_{t-1}-m_r)c_{j_4}(n_{t-1}-n_r)}{(m-m_{t-1})!(n-n_{t-1}-1)!}\int_{\xi}^{\infty}\Big(\theta_3\theta_4(1+w_t^{\theta_2})^{-\theta_1(m-m_{t-1}+j_3)}w_t^{\theta_4-1}\nonumber\\
&\times(1+w_t^{\theta_4})^{-\theta_3(n-n_{t-1}+j_4)-1}\Big)dw_t+\dfrac{c_{j_3}(m_{t-1}-m_r)c_{j_4}(n_{t-1}-n_r)}{(m-m_{t-1}-1)!(n-n_{t-1})!}\nonumber\\
&\times \int_{\xi}^{\infty}\theta_1\theta_2w_t^{\theta_2-1}(1+w_t^{\theta_2})^{-\theta_1(m-m_{t-1}+j_3)-1}
(1+w_t^{\theta_4})^{-\theta_3(n-n_{t-1}+j_4)}w_t^{\theta_2-1}dw_t\Bigg\}\nonumber\\
&\times\pi(\boldsymbol{\theta}|\textbf{w},\textbf{s})\Bigg)d\theta_1d\theta_2d\theta_3d\theta_4
\end{align}

The $100(1 - \alpha )\% $ CrIs of $W_{t}$ can be obtained by solving from the following two equations 
\begin{eqnarray*}
\bar{F}^*(w_t>L_{w_{t}}|\textbf{w,s})=
\begin{cases}
\bar{F}_{1}^{*}(w_t>L_{w_{t}}|\textbf{w},\textbf{s})=1-\frac{\alpha}{2},\quad m_r<m,\quad n_r=n,\\
\bar{F}_{2}^{*}(w_t>L_{w_{t}}|\textbf{w},\textbf{s})=1-\frac{\alpha}{2},\quad m_r=m,\quad n_r<n,\\
\bar{F}_{3}^{*}(w_t>L_{w_{t}}|\textbf{w},\textbf{s})=1-\frac{\alpha}{2},\quad m_r<m,\quad n_r<n,
\end{cases}
\end{eqnarray*}
and
\begin{eqnarray*}
\bar{F}^*(w_t>U_{W_{t}}|\textbf{w}, \textbf{s})=
\begin{cases}
\bar{F}_{1}^{*}(w_t>U_{w_{t}}|\textbf{w}, \textbf{s})=\frac{\alpha}{2},\quad m_r<m,\quad n_r=n,\\
\bar{F}_{2}^{*}(w_t>U_{w_{t}}|\textbf{w}, \textbf{s})=\frac{\alpha}{2},\quad m_r=m,\quad n_r<n,\\
\bar{F}_{3}^{*}(w_t>U_{w_{t}}|\textbf{w}, \textbf{s})=\frac{\alpha}{2},\quad m_r<m,\quad n_r<n,
\end{cases}
\end{eqnarray*}
where $\bar{F}^*(w_t>L_{w_{t}}|\textbf{w},\textbf{s})$ and $\bar{F}^*(w_t>U_{W_{t}}|\textbf{w},\textbf{s})$ are given by  \autoref{B.P.S.F}. To find HPD intervals of $W_{t}$, it is necessary to solve the following two equations simultaneously
\begin{align*}
\bar{F}^*(w_t>L_{w_{t}}|\textbf{w,s})&-\bar{F}^*(w_t>U_{W_{t}}|\textbf{w}, \textbf{s})=\\
&\begin{cases}
\bar{F}_{1}^{*}(w_t>L_{w_{t}}|\textbf{w},\textbf{s})-\bar{F}_{1}^{*}(w_t>U_{w_{t}}|\textbf{w}, \textbf{s})=1-\alpha,\quad m_r<m,\quad n_r=n,\\
\bar{F}_{2}^{*}(w_t>L_{w_{t}}|\textbf{w},\textbf{s})-\bar{F}_{2}^{*}(w_t>U_{w_{t}}|\textbf{w}, \textbf{s})=1-\alpha,\quad m_r=m,\quad n_r<n,\\
\bar{F}_{3}^{*}(w_t>L_{w_{t}}|\textbf{w},\textbf{s})-\bar{F}_{3}^{*}(w_t>U_{w_{t}}|\textbf{w}, \textbf{s})=1-\alpha,\quad m_r<m,\quad n_r<n,
\end{cases}
\end{align*}
and 
\begin{eqnarray}
f^{*}(L_{w_{t}}|\textbf{w},\textbf{s})=f^{*}(U_{w_{t}}|\textbf{w},\textbf{s})= 
\begin{cases}
f_{1}^{*}(L_{w_{t}}|\textbf{w},\textbf{s})=f_{1}^{*}(U_{w_{t}}|\textbf{w},\textbf{s}),\quad{m_r} < m,\quad{n_r} = n,\\
f_{2}^{*}(L_{w_{t}}|\textbf{w},\textbf{s})=f_{2}^{*}(U_{w_{t}}|\textbf{w},\textbf{s}),\quad{m_r} = m,\quad{n_r} < n,\\
f_{3}^{*}(L_{w_{t}}|\textbf{w},\textbf{s})=f_{3}^{*}(U_{w_{t}}|\textbf{w},\textbf{s}),\quad{m_r} < m,\quad{n_r} < n,
 \end{cases}
\end{eqnarray}
where $f^{*}(L_{w_{t}}|\textbf{w},\textbf{s})$ and $f^{*}(U_{w_{t}}|\textbf{w},\textbf{s})$ are given in \autoref{bayse prediction:w_{r+j}}.
 
\section{Simulation Study}\label{sec5}
To assess the various estimation methods discussed in this paper, a Monte Carlo simulation study was conducted using the R program (\cite{R Core Team2020}) and "bayestestR" package was used to find the CrIs and HPD intervals. The simulation was repeated $10^4$ times (represented as $n_s = 10^4$). The data was generated from the $\text{Burr-XII} (x,1.5,1)$ and  $\text{Burr-XII} (y,2,0.5)$ distributions, considering different joint type II censoring schemes and sample sizes. The specific combinations of sample sizes are as follows: $(m, n, r)$ =$\{(20, 20, 20),(20, 20, 25),(20, 20, 30), (30, 40, 35)$, 
 \\$(30, 40, 50), (30, 40, 60), (40, 30, 35),(40, 30, 50),(40, 30, 60), (50, 50, 70),(50, 50, 80), (50, 50, 85)\}$.\\ 
 Here, $m$ and $n$ represent the sizes of the first and second samples respectively, while $r$ denotes the censoring point. The first step in the analysis involved estimating the MLEs for the parameters using the EM algorithm. Subsequently, AICs were computed at the 95\% confidence level for these parameters. Additionally, Bayesian estimations of the parameters were conducted using importance sampling, considering both NIN and informative (IN) gamma prior distributions for the SE, LINEX $(v=-0.25 \text{ and } 0.5)$, and $GE (k=-0.25  \text{ and } 0.5)$ loss functions. We considered the values of hyperparameters, $(a_1=3, b_1=2)$, $(c_1=3, d_1=3)$, $(a_2=2, b_2=1)$, and $(c_2=3, d_2=6)$ for gamma prior distributions. These values were chosen such that they satisfied the conditions $E(\theta_1)=\dfrac{a_1}{b_1}$, $E(\theta_2)=\dfrac{c_1}{d_1}$, $E(\theta_3)=\dfrac{a_2}{b_2}$, and $E(\theta_4)=\dfrac{c_2}{d_2}$, respectively. The evaluation and comparison of the EM and Bayesian estimation methods were  conducted using a range of metrics, such as bias, mean square error (MSE), lower and upper limits of parameters (L and U, respectively), and interval length (IL). The bias of the estimated parameters is determined as $\text{biase}(\widehat{\theta}_i)=\widehat{\theta}_i-{\theta_i}$ for $i=1,2,3,4$. In the case of the EM and Bayesian estimations based on the SE loss function, the MSE is defined as 
\begin{flalign*}
 \text{MSE}(\widehat{\theta}_i) =\dfrac{1}{n_s}\sum_{j=1}^{n_s} (\widehat{\theta}_{i.j}-\theta_i)^{2},\quad i=1,2,3,4,
\end{flalign*}
where $\widehat{\theta}_{i.j}$ is the EM or Bayesian estimators in the ${j^{th}}$ replication for parameter $\theta_{i}$. Successively, for the Bayesian estimators obtained under the LINEX and GE loss functions, the MSE takes the following forms
\begin{equation*}
\text{MSE}_\text{{LINEX}}(\widehat{\theta}_i ) = \dfrac{1}{n_s}\sum_{j=1}^{n_s}\Big(e^{v(\widehat{\theta}_{i.j}-\theta_i)}-v(\widehat{\theta}_{i.j}-\theta_i)-1\Big),\quad i=1,2,3,4
\end{equation*}
\begin{equation*}
\text{MSE} _\text{{GE}}(\widehat{\theta}_i )=\dfrac{1}{n_s}\sum_{j=1}^{n_s}\Big(\big(\dfrac{\widehat{\theta}_{i.j}}{\theta_i}\big)^{k}-k\ln\big(\dfrac{\widehat{\theta}_{i.j}}{\theta_i}\big)-1\Big), \quad i=1,2,3,4,
\end{equation*}
where, in the $j^{th}$ replication, $\widehat{\theta}_{i.j}$ represents either the LINEX or GE Bayesian estimators for parameter $\theta_{i}$. The EM and Bayesian shrinkage estimates were calculated based on the generated data. In this case, we considered $\theta_{1.0}=1.45, \theta_{2.0}=0.99, \theta_{3.0}=1.95, \theta_{4.0}=0.45$, $w=0.5$, and $\alpha=0.05$. To compare the shrinkage estimates with the EM and Bayesian estimates, we define the RE as 
\begin{equation*}
\text{RE}(\widetilde{\theta}_i:\widehat{\theta}_i)=\frac{\text{RMSE}(\widehat{\theta}_i)}{\text{RMSE}(\widetilde{\theta}_i)},\quad i=1,2,3,4,
\end{equation*}
 where $\widetilde{\theta}_i$ represents the shrinkage estimate, $\widehat{\theta}_i$ represents the EM or Bayesian estimates, and RMSE is the root of MSE. A RE value greater than one, indicates that the estimate $\widetilde{\theta}_i$ is considered superior or more efficient than the $\widehat{\theta}_i$ estimate. Based on the simulated data, we proceed to make predictions for the first and second censored data. These predictions were made using the BUP and Bayesian prediction methods, considering both NIN and IN prior distributions. Furthermore, we calculated 95\% equal-tailed classical, CrI and HPD prediction intervals, which correspond to the L, U, and  IL. Based on the simulation results presented in Appendix A  (\autorefs{EM:sim}-\ref{P.I.E:sim}), the following results are obtained
 
\begin{itemize}

\item Based on \autoref{EM:sim}, it is evident that the results of EM estimations in terms of bias and MSE are well implemented and acceptable. In general, as the values of $m$, $n$, and $r$ increase, the bias and MSE tend to decrease.
\item Based on the results presented in \autorefs{SEL.IN:sim}, \ref{LINEX.IN:sim}, and \ref{GEL.IN:sim}, it can be concluded that the IN Bayesian estimates outperform the EM estimates in terms of bias and MSE. Furthermore, the MSEs for the LINEX and GE Bayesian estimates are smaller compared to those obtained using the SE loss function. As the values of $m$, $n$, and $r$ increase, the MSEs of the SE and LINEX Bayesian estimates decrease. For the LINEX Bayesian estimates, the MSEs are smaller for $v=-0.25$ compared to $v=0.5$, while for the GE loss function estimates, both MSEs and bias are smaller for $k=-0.25$ compared to $k=0.5$. As the values of $m$, $n$, and $r$ increase, both the bias and MSE tend to decrease for the estimates under the GE loss function. 

\item The \autorefs{SEL.NON:sim}, \ref{LINEX.NON:sim}, and \ref{GEL.NON:sim} show that the bias and MSEs for the NIN Bayesian estimates are generally lower than those for the EM estimates. Additionally, the MSEs for NIN Bayesian estimates under the SE, LINEX, and GE loss functions are higher than the MSEs for the IN Bayesian estimates under the same loss functions, respectively. Furthermore, the results show that the MSEs for LINEX and GE Bayesian estimates are lower than those for the SE Bayesian estimates, specifically, the MSEs for the LINEX Bayesian estimates for $v=-0.25$ are lower than those for $v=0.5$, and the MSEs for GE Bayesian estimates for $k=-0.25$ are lower than those for $k=0.5$. Finally, the results suggest that as the values of $m$, $n$, and $r$ increase, MSEs generally decrease for NIN Bayesian estimates. 

\item Based on the information provided in \autorefs{LS.EM:sim}-\ref{LS.GE.NIN:sem}, it is evident that all LS estimators outperform EM and Bayesian estimates in terms of RE. Generally, for the LINEX Bayesian estimates, the bias values for $V=-0.25$ are lower, while the RE values are higher compared to the bias and RE values for $V=0.5$. The same pattern holds for the GE Bayesian estimates. Additionally, the RE tends to increase as the value of $r$ increases. Furthermore, from \autorefs{SPT.EM:sim}-\ref{SPT.GEL.NON:sem} the RE values of SP estimators, is generally superior to that of EM and Bayesian estimates.  Notably, the RE values of SP estimates for IN Bayesian estimates surpasses that of NIN Bayesian estimates.

\item Based on the results presented in \autoref{L.U.CP:sim}, it can be concluded that as the values of $m$, $n$, and $r$ increase, the ILs decrease. Additionally, the ILs for CrIs and HPD intervals of IN Bayesian estimates are generally smaller compared to those of the other estimation methods. Moreover, the ILs for HPD intervals of Bayesian estimates are typically smaller than the ILs of other types of intervals.

\item The findings from \autoref{P.E:sim} indicate that the prediction values obtained from all three methods are generally close to each other. Moreover, the prediction values according to the LINEX and GE Bayesian estimates, for $v=0.5$ and $k=0.5$, respectively, are smaller compared to those for $v=-0.25$ and $k=-0.25$. Additionally, the prediction values according to IN Bayesian estimates under the GE loss function are lower than the prediction values according to IN Bayesian estimates under both the SE and LINEX loss functions
and are lower than the prediction values according to NIN Bayesian estimates under the GE loss function.
 classical predictions are generally lower than Bayesian predictions under the SE and LINEX ($v=-0.25$) loss functions. The data in \autoref{P.I.E:sim} indicates that the ILs of HPD intervals are generally lower than other intervals, particularly for IN Bayesian predictions. In addition, the ILs of CrIs for IN Bayesian predictions tend to be lower than those for NIN Bayesian predictions. As the $m$ and $n$ values increase, the ILs of CrIs decrease compared to the classical intervals. As expected, the width of prediction intervals increases for $j=2$. 
 \end{itemize}

\section{Application to Real Data}\label{sec6}
In this section, two real datasets to examine the practical application of the proposed methods are presented. These datasets were provided by \cite{Nelson1982} (Groups 1 and 6 in Table 4.1, page 462). The datasets represent the time to break down in minutes of an insulating fluid under high voltage stress. Group $3$ and $5$ are considered as the $X$ and the $Y$ samples, respectively. The details of the datasets are summarized in \autoref{realdata}. The Kolmogorov-Smirnov (K–S) test was employed to assess whether the two datasets follow the Burr-XII distribution using the "fitdistrplus" package in R software. The test results in \autoref{KS} are showed the Burr-XII distribution provides a reasonable fit for both datasets, as indicated by the p-values.
 \autoref{cessored data:real} shows the observed data under jointly type-II censored scheme with $r$=5 and 10 obtained from the two samples, presented in \autoref{realdata}. Based on \autoref{cessored data:real}, it is evident that $m_r < m$ and $n_r < n$. Therefore, it is not possible to definitively attribute the future failure to either group X or group Y. \autoref{estimate:real}, presents parameter estimations derived from EM and Bayesian methods. The hyperparameter values for the gamma prior distributions used in the Bayesian estimates are as follows: $a_1 =3$, $b_1 =4.9735$, $c_1 = 3$, $d_1 =1.003$, $a_2 = 3$, $b_2 = 5.1813$, $c_2 = 2$, and $d_2 =1.0861$. These specific values were selected to set the prior means to be approximately equal to the estimated MLEs of the parameters based on \autoref{KS}.  
\normalsize
\begin{table}[!ht]
\tabcolsep 12.5pt
\scriptsize
\centering
\caption{\small{Times to insulating fluid breakdown for groups of $X$ and $Y$.}}
\label{realdata}
\footnotesize
\begin{tabular}{ccccccccccc}
\toprule
Group&&&&&&Data\\
\toprule
 X&1.99 & 0.64 & 2.15 & 1.08 & 2.57 & 0.93  & 4.75 & 0.82 &2.06&0.49\\
Y&8.11&3.17&5.55&0.80&0.20&1.13&6.63&1.08&2.44&0.78\\
\bottomrule
\end{tabular}
\end{table}
\normalsize
\begin{table}[!ht]
\tabcolsep 28.5pt
\scriptsize
\centering
\caption{\small{MLEs and K-S test results for real datasets.}}
\label{KS}
\footnotesize
\begin{tabular}{ccccc}
\toprule
Data & && K-S distance & p-value \\
\toprule
  X &$\hat\theta_1$=0.6032&$\hat\theta_2$=2.9909 & 0.2312 &0.5819 \\
  Y &$\hat\theta_3$=0.5790& $\hat\theta_4$=1.8414 & 0.1512 & 0.9508 \\
 \bottomrule
\end{tabular}
\end{table}

\normalsize
\begin{table}[!ht]
\tabcolsep 12pt
\scriptsize
\centering
\caption{\small{Observed data from jointly type-II censored scheme with r=5 and 10 for real datasets.}}
\label{cessored data:real}
\footnotesize
\begin{tabular}{ccccccccccccccccc}
\toprule
  5 &w& 0.20& 0.49& 0.64& 0.78& 0.80\\
     &$s$&0 &1 &1 &0& 0\\

  10&w & 0.20 &0.49 &0.64& 0.78& 0.80& 0.82& 0.93& 1.08 &1.08& 1.13\\
     &$s$&0& 1& 1& 0 &0 &1& 1&1 &1& 0\\
    \bottomrule
\end{tabular}
\end{table}   
 \begin{table}[!ht]
\tabcolsep 3.2pt
\scriptsize
\begin{center}
\caption{\small{The EM and Bayesian estimates for real datasets parameters.}}
\label{estimate:real}
\begin{tabular}{ccccccccccccccccc}
\toprule
&&&\multicolumn{1}{c}{MLE}&&\multicolumn{3}{c}{NIN Bayesian estimate }&&&\multicolumn{5}{c}{IN Bayesian estimate}\\
\cmidrule(r){4-4}\cmidrule(r){5-9}\cmidrule(r){11-15}
&&&\multicolumn{1}{c}{EM}&\multicolumn{1}{c}{SE}&\multicolumn{2}{c}{LINEX}&\multicolumn{2}{c}{GE}&&\multicolumn{1}{c}{SE}&\multicolumn{2}{c}{LINEX}&\multicolumn{2}{c}{GE}\\
\cmidrule(r){6-7}\cmidrule(r){8-9}\cmidrule(r){12-13}\cmidrule(r){14-15}
$(m,n)$&r&&&&$v=-0.25$&$v$=0.5&$k=-0.25$&$k=0.5$&&&$v=-0.25$&$v_2$=0.5&$k=-0.25$&$k_2=0.5$\\
\hline

$(10,10)$&5&$\widehat\theta_1$&0.2502&0.5473&0.5584&0.5254&0.4764&0.4000&&0.5803&0.5844& 0.5721&0.5565&0.5301\\ 
&&$\widehat\theta_2$& 1.8416 & 3.4154&3.7416&2.8929&3.1323&2.8149&&3.1230&3.2807&2.8211&2.9567& 2.7703 \\ 
&&$\widehat\theta_3$&1.2143 &0.8961& 0.9099&0.8670&0.8333&0.7531&&0.6796&0.6851&0.6683&0.6496&0.6121 \\
&&$\widehat\theta_4$&3.2679 &3.0659&3.3490&2.5560&2.7550&2.4061&&2.3163&2.4637&2.0471&2.1178&1.9161\\[1pt]

&10&$\widehat\theta_1$&0.6542&0.8943&0.9066&0.8700&0.8501&0.8026&&0.7948&0.8006&0.7834& 0.7729&0.7504 \\
&&$\widehat\theta_2$&3.7840 & 4.5779&4.9731&4.0002&4.3542&4.1344&&3.7271&3.9195&3.3943&3.5774&3.4232\\
&&$\widehat\theta_3$&0.8250 &0.6362&0.6442&0.6206&0.5975&0.5571&&0.5260 &0.5320&0.5144&0.4941&0.4635\\ 
&&$\widehat\theta_4$&2.4788 & 1.8218&1.9386&1.6385&1.6585&1.5011&&2.2062&2.2704&2.0829&2.1141&2.0151\\  

 \bottomrule
 \end{tabular}
\end{center}
\end{table}
\begin{table}[!ht]
\tabcolsep 3.5pt
\scriptsize
\begin{center}
\caption{\small{95\% lower and upper bound estimates and ILs for real datasets parameters.}}
\label{L.U.CP:realdata}
\begin{tabular}{cccccccccccccccccccc}
\toprule
&&&&\multicolumn{3}{c}{${\theta_1}$}&&\multicolumn{3}{c}{${\theta_2}$}&&\multicolumn{3}{c}{${\theta_3}$}&&\multicolumn{3}{c}{${\theta_4}$}\\
\cmidrule{5-7}\cmidrule{9-11}\cmidrule{13-15}\cmidrule{17-19}

$(m,n)$&r&&&L&U&IL&&L&U&IL&&L&U&IL&&L&U&IL\\
\hline
$(10,10)$&5&&ACI& 0.0000&0.6043&0.6043&&0.0000&4.1180&4.1180&&0.0000&3.1141&3.1141&&0.0000&7.2777&7.2777\\
  &&NIN&CrI&0.0527&1.5532&1.5005&&0.2144&4.9873&4.7729&&0.1242&1.5441&1.4200&&0.2218&3.4698&3.2480 \\ 
  &&&HPD&0.0020&1.1572&1.1552&&0.0726&4.1468&4.0742&&0.0528&1.3104&1.2576&&0.0891&2.8857&2.7966 \\
           &&IN&CrI& 0.3915&0.8576&0.4661&&1.9567&3.9598&2.0031&&0.3810&0.8072&0.4262&&1.2476 & 2.9425 & 1.6949 \\ 
     &&&HPD& 0.3737&0.8309&0.4573&&1.9136&3.8935&1.9800&&0.3632&0.7885&0.4253&&1.2140 & 2.8720 & 1.6580 \\ [1pt]
     
&10&&ACI&0.1296&1.1788&1.0492&&1.1173&6.4507&5.3334&&0.0090&1.6410&1.6320&&0.2765&4.6812&4.4047\\ 
   &&NIN&CrI&0.3444&1.6673&1.3229&&1.6347&9.2335&7.5989&&0.1571&1.2833&1.1262&&0.5023&4.7228&4.2205 \\ 
  &&&HPD&0.2888&1.5184&1.2296&&1.3420&8.5341&7.1921&&0.0960&1.1060&1.0099&&0.1954&4.1942&3.9988 \\ 
           &&IN&CrI&0.4374&0.8784&0.4410&&2.2089&4.2048&1.9959&&0.3723&0.7860&0.4137&&1.1727 & 2.6549 & 1.4821 \\ 
     &&&HPD&0.4334&0.8725&0.4391&&2.1733&4.1506&1.9772&&0.3695&0.7834&0.4139&&1.1383&2.6037&1.4654 \\  

\bottomrule
 \end{tabular}
\end{center}
\end{table}
\begin{table}[!ht]
\tabcolsep 3.8pt
\scriptsize
\begin{center}
\caption{\small{LS estimator for real datasets parameters.}}
\label{LS:real}
\begin{tabular}{ccccccccccccccccc}
\toprule
&&&\multicolumn{1}{c}{MLE}&&\multicolumn{3}{c}{NIN Bayesian estimate}&&&\multicolumn{5}{c}{IN Bayesian estimate}\\
\cmidrule(r){4-4}\cmidrule(r){5-9}\cmidrule(r){11-15}
&&&\multicolumn{1}{c}{EM}&\multicolumn{1}{c}{SE}&\multicolumn{2}{c}{LINEX}&\multicolumn{2}{c}{GE}&&\multicolumn{1}{c}{SE}&\multicolumn{2}{c}{LINEX}&\multicolumn{2}{c}{GE}\\
\cmidrule(r){6-7}\cmidrule(r){8-9}\cmidrule(r){12-13}\cmidrule(r){14-15}
$(m,n)$&r&&&&$v=-0.25$&$v=0.5$&$k=-0.25$&$k=0.5$&&&$v=-0.25$&$v=0.5$&$k=-0.25$&$k=0.5$\\
\hline
$(10,10)$&5&$\widehat\theta_1$&0.4251&0.5736&0.5792& 0.5627&0.5382&0.5000&&0.5902&0.5922&0.5860&0.5783&0.5651\\
&&$\widehat\theta_2$& 2.4208&3.1077& 3.2708&2.8465& 2.9662&2.8074&&2.9615& 3.0403&2.8106&2.8783&2.7852\\
&&$\widehat\theta_3$&0.8922&0.7330&0.7400&0.7185&0.7016&0.6616&&0.6248& 0.6276& 0.6192&0.6098&0.5911\\
&&$\widehat\theta_4$&2.5839&2.4330& 2.5745&2.1780&2.2775 &2.1030&&2.0581&2.1318&1.9235&1.9589&1.8580 \\[1pt]

&10&$\widehat\theta_1$&0.6271&0.7471&0.7533&0.7350& 0.7250&0.7013&&0.6974&0.7003& 0.6917& 0.6864&0.6752\\
&&$\widehat\theta_2$& 3.3920&3.6890&3.8866&3.4001&3.5771&3.4672&&3.2636&3.3598&3.0972&3.1887&3.1116\\
&&$\widehat\theta_3$& 0.6975&0.6031& 0.6071&0.5953& 0.5838&0.5636&&0.5480&0.5510&0.5422& 0.5321&0.5168\\ 
&&$\widehat\theta_4$&2.1894&1.8109&1.8693&1.7192&1.7293&1.6506&&2.0031& 2.0352&1.9415&1.9571&1.9075\\  

 \bottomrule
 \end{tabular}
\end{center}
\end{table}
\begin{table}[!ht]
\tabcolsep 3.5pt
\scriptsize
\begin{center}
\caption{\small{SP estimator for real datasets parameters.}}
\label{SPT:real}
\begin{tabular}{ccccccccccccccccc}
\toprule
&&&\multicolumn{1}{c}{MLE}&&\multicolumn{3}{c}{NIN Bayesian estimate}&&&\multicolumn{5}{c}{IN Bayesian estimate}\\
\cmidrule(r){4-4}\cmidrule(r){5-9}\cmidrule(r){11-15}
&&&\multicolumn{1}{c}{EM}&\multicolumn{1}{c}{SE}&\multicolumn{2}{c}{LINEX}&\multicolumn{2}{c}{GE}&&\multicolumn{1}{c}{SE}&\multicolumn{2}{c}{LINEX}&\multicolumn{2}{c}{GE}\\
\cmidrule(r){6-7}\cmidrule(r){8-9}\cmidrule(r){12-13}\cmidrule(r){14-15}
$(m,n)$&r&&&&$v=-0.25$&$v=0.5$&$k=-0.25$&$k=0.5$&&&$v=-0.25$&$v=0.5$&$k=-0.25$&$k=0.5$\\
\hline
$(10,10)$&5&$\widehat\theta_1$&0.2502&0.5736&0.5792&0.5627&0.5382&0.5000&&0.5902&0.5920&0.5860& 0.5922&0.5860  &\\ 
&&$\widehat\theta_2$&1.8416&3.1077&3.2708&2.8465&2.9662&2.8074&& 2.9615&3.0403&2.8106& 3.0403& 2.8106\\
&&$\widehat\theta_3$&0.8922&0.7330&0.7400&0.7185&0.7016&0.6616&& 0.6248 &0.6276&0.6192&0.6276&0.6192 \\
&&$\widehat\theta_4$& 2.5839&2.4330&2.5745& 2.1780&2.2775&2.1030&& 2.0581 & 2.1318&1.9235&2.1318&1.9235    \\[1pt]

&10&$\widehat\theta_1$&0.6271& 0.8943&0.9066&0.7350&0.8501&0.7013&&0.6974&0.7003&0.6917&0.7003&0.6917\\
&&$\widehat\theta_2$&3.3920&4.5779& 4.9731&4.0002 &4.3542&4.1344&&3.7271&  3.9195&3.3943& 3.9195&3.3943\\
&&$\widehat\theta_3$&0.6975&0.6031& 0.6071&0.5953 &0.5838&0.5636&&0.5480& 0.5510& 0.5422& 0.5510&0.5422\\ 
&&$\widehat\theta_4$& 2.1894& 1.8109& 1.8693&1.7192&1.7293&1.6506&&2.0031 &2.0352&1.9415& 2.0352&1.9415\\  

 \bottomrule
 \end{tabular}
\end{center}
\end{table}

\begin{table}[!ht]
\tabcolsep 2.2pt
\scriptsize
\begin{center}
\caption{\small{Predictive estimate for $w_t=w_{r+j}$ based on real datasets and for $j=1,2$.}}
\label{BUP:real}
\begin{tabular}{cccccccccccccccccc}
\toprule
&&&&\multicolumn{1}{c}{Classic}&&\multicolumn{3}{c}{NIN Bayesian estimate}&&&\multicolumn{5}{c}{IN Bayesian estimate}\\
\cmidrule(r){5-5}\cmidrule(r){6-10}\cmidrule(r){12-16}
&&&&\multicolumn{1}{c}{BUP}&\multicolumn{1}{c}{SE}&\multicolumn{2}{c}{LINEX}&\multicolumn{2}{c}{GE}&&\multicolumn{1}{c}{SE}&\multicolumn{2}{c}{LINEX}&\multicolumn{2}{c}{GE}\\
\cmidrule(r){7-8}\cmidrule(r){9-10}\cmidrule(r){13-14}\cmidrule(r){15-16}
$(m,n)$&r&j&True value&&&$v=-0.25$&$v=0.5$&$k=-0.25$&$k=0.5$&&&$v=-0.25$&$v=0.5$&$k=-0.25$&$k=0.5$\\
\hline
$(10,10)$&5&1&0.82&0.8667&0.9682&0.8686&0.8960&0.9009&1.0885&&0.9008&0.8825& 0.8969&0.8982& 0.9127\\  
&&2&0.93&0.9377&1.15487&0.9375&0.9966&1.0054&1.4511&&0.9990&0.9666&0.9978&0.9997&1.0306\\

&10&1&1.99&1.2155&1.1954&1.2037&1.2287&1.2253&1.3693&&1.2153&1.2235&1.2429& 1.2405&1.2533\\
&&2&2.06&1.3110&1.2795&1.2897&1.3450&1.3379&1.6616&&1.3136&1.3315&1.3745&1.3695&1.3984\\
 \bottomrule
 \end{tabular}
\end{center}
\end{table}
\begin{table}[!ht]
\tabcolsep 1pt
\scriptsize
\begin{center}
\caption{Predictive interval estimates for $w_t=w_{r+j}$ based on real datasets and for $j=1,2$.}
\label{P.I.E:real}
\begin{tabular}{cccccccccccccccccccccccc}
\toprule

&&&&&\multicolumn{3}{c}{Classical prediction}&&\multicolumn{7}{c}{NIN Bayesian prediction}&&\multicolumn{7}{c}{IN Bayesian prediction}\\
\cmidrule(r){6-8}\cmidrule(r){10-16}\cmidrule(r){18-24}
&&&&&&&&&\multicolumn{3}{c}{CrI}&&\multicolumn{3}{c}{HPD}&&\multicolumn{3}{c}{CrI}&&\multicolumn{3}{c}{HPD}\\
\cmidrule(r){6-8}\cmidrule(r){10-12}\cmidrule(r){14-16}\cmidrule(r){18-20}\cmidrule(r){22-24}
$(m,n)$&r&&j&True value&$L_{w_{t}}$&$U_{w_{t}}$&\text{IL}&&$L_{w_{t}}$&$U_{w_{t}}$&\text{IL}&&$L_{w_{t}}$&$U_{w_{t}}$&\text{IL}&&$L_{w_{t}}$&$U_{w_{t}}$&\text{IL}&&$L_{w_{t}}$&$U_{w_{t}}$&\text{IL}\\
\hline
$(10,10)$&5&&1&0.82&0.8027&1.1610&0.3586&&0.8026&1.1450&0.3424&&0.8018&1.0880&0.2862&&0.8023&1.0973&0.2950&&0.8021&1.0710&0.2689\\
&&&2&0.93&0.7770&1.3621&0.5851&&0.8261&1.3297&0.5036&&0.8118&1.2388&0.4271&&0.8228&1.2525&0.4297&&0.8207&1.2060&0.3853\\

&10&&1&1.99&1.1339&2.0001&0.8662&&1.1320&2.0191&0.8871&&1.1317&2.0004&0.8687&&1.1328&2.0110&0.8782&&1.1354&1.9998&0.8642\\
&&&2&2.06&1.0960&2.1283&1.0323&&1.1512&2.1736&1.0224&&1.1499&2.0920&0.9421&&1.1699&2.2010&1.0311&&1.1540&2.1370&0.9830\\
 \bottomrule
\end{tabular} 
\end{center}
\end{table}
According to \autoref{estimate:real}, it can be observed that the Bayesian and EM estimations generally yield acceptable results, with the IN Bayesian estimations outperforming both the EM and NIN Bayesian estimations. Among the IN Bayesian estimations, the GE estimators demonstrate superior performance compared to the LINEX and SE estimations. Notably, as the value of $r$ increases, there is an improvement in the results for all estimators. Furthermore, from \autoref{L.U.CP:realdata}, it can be concluded that the ILs of the IN Bayesian estimations are generally smaller than the ILs of the NIN Bayesian estimations. These ILs are also lower than those obtained from the classical method. Moreover, in general, the ILs of the HPD intervals are smaller than those of other methods. From  \autoref{LS:real}, it can be seen that the LS estimates perform better than the SP estimates, and from \autoref{SPT:real}, it is evident that the SP estimates outperform both the EM and Bayesian estimations. Referring to \autoref{BUP:real}, it can be concluded that for $r=5$, classical estimates are superior to the Bayesian estimates, and generally, IN Bayesian estimates outperform the NIN Bayesian estimates.  Furthermore, \autoref{P.I.E:real} shows that the ILs of the HPD intervals, particularly for the IN Bayesian estimates, are smaller than those of other methods. Additionally, the ILs of the CrIs for the IN Bayesian estimates are also smaller than those of the NIN Bayesian estimates, and both are smaller than the ILs of the classical method. 

\section{Conclusion}\label{sec7}
In this article, we have investigated the problem of linear shrinkage and shrinkage pretest estimations and prediction for two Burr-XII distributions under the joint type-II censoring scheme. We have compared the shrinkage estimations with Bayesian and EM estimations developed by \cite{Akbari Bargoshadi2023}. Additionally, we have presented two real datasets to illustrate all the inferential results. Based on the simulation results, we can conclude that Bayesian estimations are better than EM estimations in terms of bias and MSE. However, shrinkage estimations outperform other methods in terms of RE. Moreover, as the values of $m$, $n$, and $r$ increases, the IL of HPD intervals, especially for IN Bayesian estimates, is smaller compared to the IL of intervals for other methods. Furthermore, the prediction values for the BUP are generally lower than those for other methods, and the IL of HPD prediction intervals is also lower than IL of intervals for other methods. Additionaly, the IL of CrIs for IN Bayesian estimates are typically lower than those for NIN Bayesian estimates.

\newpage
\section{Appendix A}
\label{appendix}
\setcounter{table}{0}
\renewcommand{\thetable}{A\arabic{table}}
\renewcommand*{\theHtable}{\thetable}

\normalsize
\begin{table}[!ht]
\tabcolsep 12pt
\scriptsize
\begin{center}
\caption{\small{Bias and MSE of EM estimations.}}
 \label{EM:sim}

\begin{tabular}{cccccccccc}
\toprule
&&\multicolumn{2}{c}{${\widehat\theta_1}$}&\multicolumn{2}{c}{${\widehat\theta_2}$}&\multicolumn{2}{c}{${\widehat\theta_3}$}&\multicolumn{2}{c}{${\widehat\theta_4}$}\\
\cmidrule(lr){3-4} \cmidrule(lr){5-6} \cmidrule(lr){7-8} \cmidrule(lr){9-10}
$(m,n)$&$r$&Bias&MSE&Bias&MSE&Bias&MSE&Bias&MSE\\
\hline
  
$(20,20)$&20 & 0.5027 & 1.6660 & 0.1695 & 0.1730 & 0.2578 & 0.6777 & 0.0373 & 0.0181 \\ 
              &25& 0.2836 & 0.6207 & 0.1214 & 0.1111 & 0.2475 & 0.7037 & 0.0353 & 0.0142 \\ 
              &30& 0.1575 & 0.2911 & 0.0994 & 0.0870 & 0.1733 & 0.5075 & 0.0401 & 0.0150 \\
               
$(30,40)$&35 & 0.2966 & 0.9460 & 0.0925 & 0.0858 & 0.1548 & 0.3061 & 0.0224 & 0.0090 \\  
              &50& 0.1017 & 0.1754 & 0.0617 & 0.0446 & 0.0812 & 0.1678 & 0.0189 & 0.0066 \\ 
              &60 & 0.0374 & 0.1063 & 0.0570 & 0.0364 & 0.0717 & 0.1509 & 0.0126 & 0.0052 \\  
              
$(40,30)$&35& 0.1834 & 0.4382 & 0.0771 & 0.0568 & 0.1825 & 0.4233 & 0.0251 & 0.0104 \\
              &50 & 0.0756 & 0.1278 & 0.0447 & 0.0307 & 0.1131 & 0.2516 & 0.0170 & 0.0078 \\ 
              &60& 0.0414 & 0.0733 & 0.0327 & 0.0223 & 0.0914 & 0.1970 & 0.0171 & 0.0069 \\
  
$(50,50)$&70 & 0.0636 & 0.0914 & 0.0406 & 0.0268 & 0.0613 & 0.1298 & 0.0159 & 0.0051 \\ 
              &80& 0.0423 & 0.0688 & 0.0308 & 0.0192 & 0.0562 & 0.1170 & 0.0136 & 0.0046 \\
              &85 & 0.0263 & 0.0562 & 0.0270 & 0.0186 & 0.0442 & 0.1089 & 0.0089 & 0.0040 \\ 
\bottomrule
\end{tabular}
\end{center}
\end{table}

\normalsize
\begin{table}[!ht]
\tabcolsep 12pt
\scriptsize
\begin{center}
\caption{\small{Bias and MSE of IN Bayesian estimations for SE loss function.}}
 \label{SEL.IN:sim}
\begin{tabular}{cccccccccc}
\toprule
&&\multicolumn{2}{c}{${\widehat\theta_1}$}&\multicolumn{2}{c}{${\widehat\theta_2}$}&\multicolumn{2}{c}{${\widehat\theta_3}$}&\multicolumn{2}{c}{${\widehat\theta_4}$}\\
\cmidrule(lr){3-4} \cmidrule(lr){5-6} \cmidrule(lr){7-8} \cmidrule(lr){9-10}
$(m,n)$&$r$&Bias&MSE&Bias&MSE&Bias&MSE&Bias&MSE\\
\hline 
$(20,20)$&20 & -0.0101 & 0.1534 & -0.0357 & 0.0493 & 0.0067 & 0.2856 & -0.0137 & 0.0100 \\ 
              &25 & 0.0281 & 0.1423 & 0.0137 & 0.0495 & 0.0390 & 0.2359 & -0.0015 & 0.0095 \\
             &30& 0.0679 & 0.1414 & 0.0434 & 0.0427 & 0.0599 & 0.2170 & 0.0132 & 0.0089 \\ 
  
$(30,40)$&35 & -0.1382 & 0.1690 & -0.1394 & 0.0560 & -0.1481 & 0.1572 & -0.0670 & 0.0103 \\
              &50 & 0.0320 & 0.1008 & -0.0098 & 0.0343 & -0.0001 & 0.1309 & -0.0197 & 0.0053 \\
              &60 & 0.0387 & 0.0866 & 0.0352 & 0.0312 & 0.0598 & 0.1204 & 0.0066 & 0.0048 \\ 
  
 $(40,30)$&35 & -0.1242 & 0.1105 & -0.1368 & 0.0459 & -0.1192 & 0.1926 & -0.0578 & 0.0096 \\ 
               &50& 0.0171 & 0.0776 & -0.0109 & 0.0245 & 0.0142 & 0.1731 & -0.0125 & 0.0063 \\
               &60 & 0.0181 & 0.0593 & 0.0243 & 0.0206 & 0.0479 & 0.1544 & 0.0107 & 0.0065 \\ 
  
 $(50,50)$&70 & -0.0111 & 0.0658 & -0.0533 & 0.0244 & -0.0556 & 0.0983 & -0.0348 & 0.0055 \\  
               &80 & 0.0130 & 0.0560 & 0.0051 & 0.0192 & -0.0034 & 0.0910 & -0.0042 & 0.0042 \\ 
               &85 & 0.0252 & 0.0604 & 0.0231 & 0.0168 & 0.0313 & 0.0845 & 0.0017 & 0.0040 \\ 
\bottomrule
\end{tabular}
\end{center}
\end{table}
\begin{table}[!ht]
\tabcolsep 1.1pt
\scriptsize
\begin{center}
\caption{\small{Bias and MSE of IN Bayesian estimations for LINEX  loss function.}}
 \label{LINEX.IN:sim}
\begin{tabular}{cccccccccccccccccccccccc}
\toprule
&&&&&&\multicolumn{4}{c}{$v_1=-0.25$}&&&&&&&\multicolumn{4}{c}{$v_2=0.5$}\\
\cmidrule(r){5-12} \cmidrule(r){15-22}
&&&&\multicolumn{2}{c}{${\widehat\theta_1}$}&\multicolumn{2}{c}{${\widehat\theta_2}$}&\multicolumn{2}{c}{${\widehat\theta_3}$}&\multicolumn{2}{c}{${\widehat\theta_4}$}&&&\multicolumn{2}{c}{${\widehat\theta_1}$}&\multicolumn{2}{c}{${\widehat\theta_2}$}&\multicolumn{2}{c}{${\widehat\theta_3}$}&\multicolumn{2}{c}{${\widehat\theta_4}$}\\
\cmidrule(r){5-6} \cmidrule(r){7-8} \cmidrule(r){9-10} \cmidrule(r){11-12}\cmidrule(r){15-16} \cmidrule(r){17-18} \cmidrule(r){19-20} \cmidrule(r){21-22}
$(m,n)$&&$r$&&Bias&MSE&Bias&MSE&Bias&MSE&Bias&MSE&&&Bias&MSE&Bias&MSE&Bias&MSE&Bias&MSE\\
\hline
$(20,20)$&&20&& 0.0178 & 0.0050 & -0.0303 & 0.0016 & 0.0412 & 0.0090 & -0.0128 & 0.0003 &&& -0.0624 & 0.0181 & -0.0464 & 0.0060 & -0.0597 & 0.0349 & -0.0156 & 0.0012 \\
  &&25 && 0.0498 & 0.0046 & 0.0192 & 0.0016 & 0.0717 & 0.0075 & -0.0005 & 0.0003 &&& -0.0137 & 0.0169 & 0.0029 & 0.0060 & -0.0232 & 0.0286 & -0.0035 & 0.0012 \\
  &&30&& 0.0863 & 0.0045 & 0.0488 & 0.0013 & 0.0914 & 0.0069 & 0.0142 & 0.0003 &&& 0.0326 & 0.0169 & 0.0329 & 0.0052 & 0.0004 & 0.0259 & 0.0110 & 0.0011 \\ 
  
$(30,40)$&&35&& -0.1233 & 0.0053 & -0.1369 & 0.0018 & -0.1356 & 0.0050 & -0.0668 & 0.0003 &&& -0.1671 & 0.0207 & -0.1446 & 0.0068 & -0.1729 & 0.0194 & -0.0676 & 0.0013 \\  
  &&50& & 0.0434 & 0.0032 & -0.0067 & 0.0011 & 0.0135 & 0.0041 & -0.0193 & 0.0002 &&& 0.0097 & 0.0124 & -0.0159 & 0.0042 & -0.0271 & 0.0166 & -0.0204 & 0.0007 \\
\bottomrule 
\end{tabular}
\end{center}
\end{table}
\begin{table}[!ht]
\tabcolsep 1.1pt
\scriptsize
\begin{center}
\caption{\small{Bias and MSE of IN Bayesian estimations for GE loss function.}}
 \label{GEL.IN:sim}
\begin{tabular}{cccccccccccccccccccccc}
\toprule
&&&&&&\multicolumn{4}{c}{$k_1=-0.25$}&&&&&&&\multicolumn{4}{c}{$k_2=0.5$}\\
\cmidrule(r){5-12} \cmidrule(r){15-22}
&&&&\multicolumn{2}{c}{${\widehat\theta_1}$}&\multicolumn{2}{c}{${\widehat\theta_2}$}&\multicolumn{2}{c}{${\widehat\theta_3}$}&\multicolumn{2}{c}{${\widehat\theta_4}$}&&&\multicolumn{2}{c}{${\widehat\theta_1}$}&\multicolumn{2}{c}{${\widehat\theta_2}$}&\multicolumn{2}{c}{${\widehat\theta_3}$}&\multicolumn{2}{c}{${\widehat\theta_4}$}\\
\cmidrule(r){5-6} \cmidrule(r){7-8} \cmidrule(r){9-10} \cmidrule(r){11-12}\cmidrule(r){15-16} \cmidrule(r){17-18} \cmidrule(r){19-20} \cmidrule(r){21-22}
$(m,n)$&&$r$&&Bias&MSE&Bias&MSE&Bias&MSE&Bias&MSE&&&Bias&MSE&Bias&MSE&Bias&MSE&Bias&MSE\\
\hline
$(20,20)$&& 20 && -0.0630 & 0.0024 & -0.0524 & 0.0018 & -0.0432 & 0.0023 & -0.0195 & 0.0014 &&& -0.1170 & 0.0102 & -0.0702 & 0.0073 & -0.0945 & 0.0093 & -0.0255 & 0.0055 \\ 
 &&25& & -0.0126 & 0.0019 & -0.0021 & 0.0015 & -0.0071 & 0.0017 & -0.0075 & 0.0012 &&& -0.0540 & 0.0078 & -0.0186 & 0.0059 & -0.0538 & 0.0072 & -0.0136 & 0.0047 \\ 
 &&30& & 0.0348 & 0.0017 & 0.0286 & 0.0011 & 0.0166 & 0.0015 & 0.0070 & 0.0010 &&& 0.0015 & 0.0068 & 0.0135 & 0.0045 & -0.0271 & 0.0063 & 0.0007 & 0.0041 \\ 
  
 $(30,40)$&& 35 && -0.1697 & 0.0033 & -0.1485 & 0.0026 & -0.1683 & 0.0016 & -0.0689 & 0.0018 &&& -0.2021 & 0.0129 & -0.1580 & 0.0101 & -0.1889 & 0.0063 & -0.0708 & 0.0068 \\
 &&50&& 0.0101 & 0.0013 & -0.0189 & 0.0011 & -0.0205 & 0.0010 & -0.0220 & 0.0008 &&& -0.0121 & 0.0054 & -0.0282 & 0.0043 & -0.0412 & 0.0041 & -0.0244 & 0.0030 \\
\bottomrule
\end{tabular}
\end{center}
\end{table}

\begin{table}[!ht]
\tabcolsep 12.6pt
\scriptsize
\begin{center}
\caption{\small{Bias and MSE of NIN Bayesian estimations for SE loss function.}}
 \label{SEL.NON:sim}
\begin{tabular}{cccccccccc}
\toprule
&&\multicolumn{2}{c}{${\widehat\theta_1}$}&\multicolumn{2}{c}{${\widehat\theta2}$}&\multicolumn{2}{c}{${\widehat\theta_3}$}&\multicolumn{2}{c}{${\widehat\theta_4}$}\\
\cmidrule(lr){3-4} \cmidrule(lr){5-6} \cmidrule(lr){7-8} \cmidrule(lr){9-10}
$(m,n)$&$r$&Bias&MSE&Bias&MSE&Bias&MSE&Bias&MSE\\
\hline
$(20,20)$&20 & 0.0115 & 0.4685 & -0.0564 & 0.0864 & -0.0015 & 0.4112 & -0.0259 & 0.0135 \\ 
              &25 & 0.0740 & 0.2682 & 0.0074 & 0.0768 & 0.0512 & 0.3520 & -0.0131 & 0.0116 \\ 
              &30 & 0.1107 & 0.2448 & 0.0551 & 0.0620 & 0.0729 & 0.3318 & 0.0098 & 0.0126 \\ 
  
$(30,40)$&35& -0.2518 & 0.2569 & -0.2026 & 0.0838 & -0.2052 & 0.2080 & -0.0782 & 0.0116 \\
               &50 & 0.0171 & 0.1330 & -0.0155 & 0.0431 & -0.0075 & 0.1324 & -0.0261 & 0.0061 \\
               &60 & 0.0425 & 0.1210 & 0.0691 & 0.1586 & 0.0568 & 0.1470 & 0.0043 & 0.0047 \\ 
  
$(40,30)$&35& -0.2037 & 0.1606 & -0.1860 & 0.0662 & -0.1296 & 0.2442 & -0.0719 & 0.0132 \\ 
             &50 & 0.0165 & 0.0914 & -0.0156 & 0.0320 & 0.0355 & 0.2081 & -0.0136 & 0.0081 \\ 
              &60 & 0.0371 & 0.0774 & 0.0426 & 0.0319 & 0.0526 & 0.1884 & 0.0109 & 0.0073 \\ 
  
$(50,50)$&70&  -0.0091 & 0.0807 & -0.0567 & 0.0306 & -0.0443 & 0.1094 & -0.0420 & 0.0063 \\ 
              &80 & 0.0258 & 0.0737 & 0.0083 & 0.0215 & 0.0048 & 0.1041 & -0.0110 & 0.0047 \\ 
              &85 & 0.0242 & 0.0625 & 0.0191 & 0.0194 & 0.0081 & 0.0963 & -0.0005 & 0.0042 \\ 
\bottomrule              
\end{tabular}
\end{center}
\end{table}
\begin{table}[!ht]
\tabcolsep 1.1pt
\scriptsize
\begin{center}
\caption{\small{Bias and MSE of NIN Bayesian estimations for LINEX loss function.}}
 \label{LINEX.NON:sim}
\begin{tabular}{ccccccccccccccccccccccc}
\toprule
&&&&&&\multicolumn{4}{c}{$v_1=-0.25$}&&&&&&&\multicolumn{4}{c}{$v_2=0.5$}\\
\cmidrule(r){5-12} \cmidrule(r){15-22}
&&&&\multicolumn{2}{c}{${\widehat\theta_1}$}&\multicolumn{2}{c}{${\widehat\theta_2}$}&\multicolumn{2}{c}{${\widehat\theta_3}$}&\multicolumn{2}{c}{${\widehat\theta_4}$}&&&\multicolumn{2}{c}{${\widehat\theta_1}$}&\multicolumn{2}{c}{${\widehat\theta_2}$}&\multicolumn{2}{c}{${\widehat\theta_3}$}&\multicolumn{2}{c}{${\widehat\theta_4}$}\\
\cmidrule(r){5-6} \cmidrule(r){7-8} \cmidrule(r){9-10} \cmidrule(r){11-12}\cmidrule(r){15-16} \cmidrule(r){17-18} \cmidrule(r){19-20} \cmidrule(r){21-22}
$(m,n)$&&$r$&&Bias&MSE&Bias&MSE&Bias&MSE&Bias&MSE&&&Bias&MSE&Bias&MSE&Bias&MSE&Bias&MSE\\
\hline
$(20,20)$&&20 && 0.0580 & 0.0147 & -0.0498 & 0.0027 & 0.0384 & 0.0128 & -0.0250 & 0.0004 &&& -0.0741 & 0.0641 & -0.0696 & 0.0109 & -0.0781 & 0.0516 & -0.0279 & 0.0017 \\ 
 &&25 && 0.1048 & 0.0086 & 0.0137 & 0.0024 & 0.0879 & 0.0108 & -0.0120 & 0.0004 &&& 0.0161 & 0.0318 & -0.0051 & 0.0098 & -0.0183 & 0.0468 & -0.0152 & 0.0014 \\ 
  &&30&& 0.1341 & 0.0076 & 0.0612 & 0.0019 & 0.1099 & 0.0105 & 0.0110 & 0.0004 &&& 0.0663 & 0.0307 & 0.0431 & 0.0077 & 0.0031 & 0.0398 & 0.0074 & 0.0016 \\
  
$(30,40)$&&35&& -0.2349 & 0.0081 & -0.1998 & 0.0027 & -0.1926 & 0.0065 & -0.0779 & 0.0004 &&& -0.2842 & 0.0314 & -0.2082 & 0.0101 & -0.2303 & 0.0261 & -0.0787 & 0.0014 \\
  &&50 && 0.0300 & 0.0042 & -0.0122 & 0.0014 & 0.0068 & 0.0041 & -0.0257 & 0.0002&&& -0.0081 & 0.0166 & -0.0222 & 0.0054 & -0.0354 & 0.0168 & -0.0269 & 0.0008 \\
\bottomrule
\end{tabular}
\end{center}
\end{table}

\begin{table}[!ht]
\tabcolsep 1.1pt
\scriptsize
\begin{center}
\caption{\small{Bias and MSE of NIN Bayesian estimations for GE loss function.}}
 \label{GEL.NON:sim}
\begin{tabular}{cccccccccccccccccccccc}
\toprule
&&&&&&\multicolumn{4}{c}{$k_1=-0.25$}&&&&&&&\multicolumn{4}{c}{$k_2=0.5$}\\
\cmidrule(r){5-12} \cmidrule(r){15-22}
&&&&\multicolumn{2}{c}{${\widehat\theta_1}$}&\multicolumn{2}{c}{${\widehat\theta_2}$}&\multicolumn{2}{c}{${\widehat\theta_3}$}&\multicolumn{2}{c}{${\widehat\theta_4}$}&&&\multicolumn{2}{c}{${\widehat\theta_1}$}&\multicolumn{2}{c}{${\widehat\theta_2}$}&\multicolumn{2}{c}{${\widehat\theta_3}$}&\multicolumn{2}{c}{${\widehat\theta_4}$}\\
\cmidrule(r){5-6} \cmidrule(r){7-8} \cmidrule(r){9-10} \cmidrule(r){11-12}\cmidrule(r){15-16} \cmidrule(r){17-18} \cmidrule(r){19-20} \cmidrule(r){21-22}
$(m,n)$&&$r$&&Bias&MSE&Bias&MSE&Bias&MSE&Bias&MSE&&&Bias&MSE&Bias&MSE&Bias&MSE&Bias&MSE\\
\hline
$(20,20)$&&20&& -0.0690 & 0.0065 & -0.0778 & 0.0033 & -0.0588 & 0.0033 & -0.0323 & 0.0020 &&& -0.1524 & 0.0260 & -0.1010 & 0.0132 & -0.1178 & 0.0134 & -0.0390 & 0.0080 \\ 
 &&25& & 0.0209 & 0.0032 & -0.0108 & 0.0022 & 0.0000 & 0.0024 & -0.0194 & 0.0016 &&& -0.0335 & 0.0129 & -0.0300 & 0.0087 & -0.0521 & 0.0099 & -0.0261 & 0.0063 \\ 
  &&30& & 0.0708 & 0.0025 & 0.0386 & 0.0016 & 0.0230 & 0.0023 & 0.0030 & 0.0014 &&& 0.0303 & 0.0103 & 0.0216 & 0.0063 & -0.0275 & 0.0094 & -0.0041 & 0.0058 \\ 
  
  $(30,40)$&&35&& -0.2900 & 0.0065 & -0.2133 & 0.0046 & -0.2265 & 0.0022 & -0.0802 & 0.0021 &&& -0.3292 & 0.0248 & -0.2247 & 0.0172 & -0.2484 & 0.0086 & -0.0823 & 0.0079 \\ 
  &&50 && -0.0076 & 0.0018 & -0.0255 & 0.0014 & -0.0287 & 0.0010 & -0.0285 & 0.0009 &&& -0.0327 & 0.0073 & -0.0357 & 0.0055 & -0.0502 & 0.0043 & -0.0310 & 0.0035 \\
\bottomrule   
\end{tabular}
\end{center}
\end{table}

\begin{table}[!ht]
\tabcolsep 12.5pt
\scriptsize
\begin{center}
\caption{\small{Bias and RE values of LS estimates for EM estimations.}}
 \label{LS.EM:sim}

\begin{tabular}{cccccccccc}
\toprule
&&\multicolumn{2}{c}{${\widehat\theta_1}$}&\multicolumn{2}{c}{${\widehat\theta_2}$}&\multicolumn{2}{c}{${\widehat\theta_3}$}&\multicolumn{2}{c}{${\widehat\theta_4}$}\\
\cmidrule(lr){3-4} \cmidrule(lr){5-6} \cmidrule(lr){7-8} \cmidrule(lr){9-10}
$(m,n)$&$r$&Bias&RE&Bias&RE&Bias&RE&Bias&RE\\
\hline

$(20,20)$&20& 0.2263 & 2.0293 & 0.0798 & 2.0193 & 0.1039 & 2.0353 & -0.0064 & 2.0713 \\ [1pt]
              &25 & 0.1168 & 2.0430 & 0.0557 & 2.0213 & 0.0988 & 2.0324 & -0.0073 & 2.0773 \\ [1pt]
              &30 & 0.0538 & 2.0471 & 0.0447 & 2.0221 & 0.0616 & 2.0299 & -0.0050 & 2.1088 \\ [1pt]
  
$(30,40)$&35& 0.1233 & 2.0293 & 0.0412 & 2.0207 & 0.0524 & 2.0438 & -0.0138 & 1.9720 \\ [1pt]
               &50& 0.0258 & 2.0452 & 0.0258 & 2.0259 & 0.0156 & 2.0344 & -0.0156 & 1.9132 \\ [1pt]
               &60 & -0.0063 & 2.0118 & 0.0235 & 2.0292 & 0.0108 & 2.0317 & -0.0187 & 1.7971 \\[1pt] 
  
$(40,30)$&35& 0.0667 & 2.0372 & 0.0335 & 2.0259 & 0.0663 & 2.0383 & -0.0125 & 2.0008 \\ [1pt]
              &50& 0.0128 & 2.0408 & 0.0174 & 2.0264 & 0.0316 & 2.0360 & -0.0165 & 1.9049 \\ [1pt]
              &60 & -0.0043 & 2.0228 & 0.0113 & 2.0253 & 0.0207 & 2.0346 & -0.0165 & 1.8946 \\ [1pt]
  
$(50,50)$&70& 0.0068 & 2.0436 & 0.0153 & 2.0271 & 0.0056 & 2.0285 & -0.0170 & 1.8423 \\[1pt] 
              &80& -0.0038 & 2.0256 & 0.0104 & 2.0275 & 0.0031 & 2.0272 & -0.0182 & 1.7910 \\ [1pt]
              &85 & -0.0119 & 2.0023 & 0.0085 & 2.0241 & -0.0029 & 2.0178 & -0.0206 & 1.6896 \\ 
\bottomrule
\end{tabular}
\end{center}
\end{table}
\begin{table}[!ht]
\tabcolsep 12pt
\scriptsize
\begin{center}
\caption{\small{Bias and RE values of LS estimates for IN Bayesian estimates under SE loss function.}}
 \label{LS.SEL.IN:sim}
\begin{tabular}{cccccccccc}
\toprule
&&\multicolumn{2}{c}{${\widehat\theta_1}$}&\multicolumn{2}{c}{${\widehat\theta_2}$}&\multicolumn{2}{c}{${\widehat\theta_3}$}&\multicolumn{2}{c}{${\widehat\theta_4}$}\\
\cmidrule(lr){3-4} \cmidrule(lr){5-6} \cmidrule(lr){7-8} \cmidrule(lr){9-10}
$(m,n)$&$r$&Bias&RE&Bias&RE&Bias&RE&Bias&RE\\
\hline
$(20,20)$&20& -0.0300 & 1.9775 & -0.0228 & 1.9837 & -0.0217 & 1.9936 & -0.0319 & 1.6979 \\ 
              &25 & -0.0110 & 2.0021 & 0.0019 & 2.0035 & -0.0055 & 2.0060 & -0.0258 & 1.7680 \\  
              &30 & 0.0089 & 2.0310 & 0.0167 & 2.0183 & 0.0050 & 2.0163 & -0.0184 & 1.8796 \\  
  
$(30,40)$&35 & -0.0941 & 1.9099 & -0.0747 & 1.9504 & -0.0991 & 1.8982 & -0.0585 & 1.4526 \\ 
              &50 & -0.0090 & 2.0070 & -0.0099 & 1.9914 & -0.0251 & 1.9811 & -0.0348 & 1.4742 \\ 
             &60 & -0.0056 & 2.0160 & 0.0126 & 2.0197 & 0.0049 & 2.0296 & -0.0217 & 1.7014 \\ 
  
$(40,30)$&35& -0.0871 & 1.8773 & -0.0734 & 1.9409 & -0.0846 & 1.9291 & -0.0539 & 1.4670 \\  
             &50 & -0.0164 & 1.9900 & -0.0104 & 1.9872 & -0.0179 & 1.9938 & -0.0313 & 1.5833 \\
              &60 & -0.0160 & 1.9884 & 0.0072 & 2.0190 & -0.0011 & 2.0150 & -0.0197 & 1.8114 \\  
  
$(50,50)$&70 & -0.0305 & 1.9474 & -0.0317 & 1.9538 & -0.0528 & 1.9227 & -0.0424 & 1.3832 \\ 
             &80 & -0.0185 & 1.9789 & -0.0025 & 2.0001 & -0.0267 & 1.9695 & -0.0271 & 1.5355 \\ 
             &85& -0.0124 & 2.0003 & 0.0066 & 2.0220 & -0.0093 & 2.0075 & -0.0241 & 1.5870 \\ 
  \bottomrule
\end{tabular}
\end{center}
\end{table}
\begin{table}[!ht]
\tabcolsep 1.1pt
\scriptsize
\begin{center}
\caption{\small{Bias and RE values of LS estimates for IN Bayesian estimates under LINEX loss function.}}
 \label{LS.LINEXL.IN:sem}
\begin{tabular}{cccccccccccccccccccccc}
\toprule
&&&&&&\multicolumn{4}{c}{$v_1=-0.25$}&&&&&&&\multicolumn{4}{c}{$v_2=0.5$}\\
\cmidrule(r){5-12} \cmidrule(r){15-22}
&&&&\multicolumn{2}{c}{${\widehat\theta_1}$}&\multicolumn{2}{c}{${\widehat\theta_2}$}&\multicolumn{2}{c}{${\widehat\theta_3}$}&\multicolumn{2}{c}{${\widehat\theta_4}$}&&&\multicolumn{2}{c}{${\widehat\theta_1}$}&\multicolumn{2}{c}{${\widehat\theta_2}$}&\multicolumn{2}{c}{${\widehat\theta_3}$}&\multicolumn{2}{c}{${\widehat\theta_4}$}\\
\cmidrule(r){5-6} \cmidrule(r){7-8} \cmidrule(r){9-10} \cmidrule(r){11-12}\cmidrule(r){15-16} \cmidrule(r){17-18} \cmidrule(r){19-20} \cmidrule(r){21-22}
$(m,n)$&&$r$&&Bias&RE&Bias&RE&Bias&RE&Bias&RE&&&Bias&RE&Bias&RE&Bias&RE&Bias&RE\\
\hline
$(20,20)$&&20&& -0.0161 & 1.9758 & -0.0202 & 1.9836 & -0.0044 & 1.9749 & -0.0314 & 1.6986 &&& -0.0562 & 1.9696 & -0.0282 & 1.9825 & -0.0548 & 2.0292 & -0.0328 & 1.6966 \\ 
              &&25& & -0.0001 & 1.9924 & 0.0046 & 1.9962 & 0.0109 & 1.9892 & -0.0253 & 1.7692 &&& -0.0319 & 2.0123 & -0.0036 & 2.0168 & -0.0366 & 2.0326 & -0.0268 & 1.7652 \\ 
        &&30& & 0.0181 & 2.0159 & 0.0194 & 2.0077 & 0.0207 & 1.9996 & -0.0179 & 1.8813  &&& -0.0087 & 2.0553 & 0.0115 & 2.0387 & -0.0248 & 2.0417 & -0.0195 & 1.8755 \\ 
  
$(30,40)$&&35&& -0.0867 & 1.9123 & -0.0734 & 1.9596 & -0.0928 & 1.9063 & -0.0584 & 1.4542  &&& -0.1086 & 1.9030 & -0.0773 & 1.9330 & -0.1114 & 1.8845 & -0.0588 & 1.4495 \\ 
          &&50& & -0.0033 & 1.9997 & -0.0083 & 1.9875 & -0.0182 & 1.9726 & -0.0347 & 1.4749  &&& -0.0202 & 2.0194 & -0.0130 & 1.9989 & -0.0386 & 1.9998 & -0.0352 & 1.4728 \\
\bottomrule
\end{tabular}
\end{center}
\end{table}
\begin{table}[!ht]
\tabcolsep 1.1pt
\scriptsize
\begin{center}
\caption{\small{Bias and RE values of LS estimates for IN Bayesian estimates under GE loss function.}}
 \label{LS.GEL.IN:sem}
\begin{tabular}{cccccccccccccccccccccc}
\toprule
&&&&&&\multicolumn{4}{c}{$k_1=-0.25$}&&&&&&&\multicolumn{4}{c}{$k_2=0.5$}\\
\cmidrule(r){5-12} \cmidrule(r){15-22}
&&&&\multicolumn{2}{c}{${\widehat\theta_1}$}&\multicolumn{2}{c}{${\widehat\theta_2}$}&\multicolumn{2}{c}{${\widehat\theta_3}$}&\multicolumn{2}{c}{${\widehat\theta_4}$}&&&\multicolumn{2}{c}{${\widehat\theta_1}$}&\multicolumn{2}{c}{${\widehat\theta_2}$}&\multicolumn{2}{c}{${\widehat\theta_3}$}&\multicolumn{2}{c}{${\widehat\theta_4}$}\\
\cmidrule(r){5-6} \cmidrule(r){7-8} \cmidrule(r){9-10} \cmidrule(r){11-12}\cmidrule(r){15-16} \cmidrule(r){17-18} \cmidrule(r){19-20} \cmidrule(r){21-22}
$(m,n)$&&$r$&&Bias&RE&Bias&RE&Bias&RE&Bias&RE&&&Bias&RE&Bias&RE&Bias&RE&Bias&RE\\
\hline
$(20,20)$&&20 && -0.0565 & 2.0287 & -0.0312 & 2.0950 & -0.0466 & 2.0181 & -0.0347 & 1.6152 &&&  -0.0835 & 2.0043 & -0.0401 & 2.0697 & -0.0722 & 2.0046 & -0.0377 & 1.6023 \\ 
    &&25&&-0.0313 & 1.9687 & -0.0061 & 2.0134 & -0.0285 & 1.9816 & -0.0287 & 1.6335 &&&  -0.0520 & 1.9601 & -0.0143 & 2.0108 & -0.0519 & 1.9769 & -0.0318 & 1.6198 \\ 
    &&30&& -0.0076 & 1.9533 & 0.0093 & 1.9714 & -0.0167 & 1.9724 & -0.0215 & 1.6801 &&& -0.0243 & 1.9539 & 0.0017 & 1.9808 & -0.0385 & 1.9705 & -0.0246 & 1.6642 \\
     
$(30,40)$&&35 && -0.1098 & 2.0563 & -0.0792 & 2.1637 & -0.1091 & 2.0020 & -0.0594 & 1.5304 &&&  -0.1260 & 2.0156 & -0.0840 & 2.1112 & -0.1194 & 1.9711 & -0.0604 & 1.5123 \\ 
    &&50 && -0.0199 & 1.9504 & -0.0144 & 2.0073 & -0.0353 & 1.9461 & -0.0360 & 1.4402 &&&  -0.0310 & 1.9442 & -0.0191 & 2.0005 & -0.0456 & 1.9405 & -0.0372 & 1.4328 \\
\bottomrule
\end{tabular}
\end{center}
\end{table}
\begin{table}[!ht]
\tabcolsep 12pt
\scriptsize
\begin{center}
\caption{\small{Bias and RE values of LS estimates for NIN Bayesian estimates under SE loss function.}}
 \label{LS.SEL.NIN:sim}
\begin{tabular}{cccccccccc}
\toprule
&&\multicolumn{2}{c}{${\widehat\theta_1}$}&\multicolumn{2}{c}{${\widehat\theta_2}$}&\multicolumn{2}{c}{${\widehat\theta_3}$}&\multicolumn{2}{c}{${\widehat\theta_4}$}\\
\cmidrule(lr){3-4} \cmidrule(lr){5-6} \cmidrule(lr){7-8} \cmidrule(lr){9-10}
$(m,n)$&$r$&Bias&RE&Bias&RE&Bias&RE&Bias&RE\\
\hline

$(20,20)$&20& -0.0193 & 1.9971 & -0.0332 & 1.9859 & -0.0257 & 1.9936 & -0.0380 & 1.7038 \\ 
              &25 & 0.0120 & 2.0185 & -0.0013 & 2.0006 & 0.0006 & 2.0075 & -0.0315 & 1.7358 \\ 
             &30 & 0.0304 & 2.0360 & 0.0226 & 2.0164 & 0.0115 & 2.0146 & -0.0201 & 1.8892 \\ 
  
$(30,40)$&35 & -0.1509 & 1.9002 & -0.1063 & 1.9522 & -0.1276 & 1.8978 & -0.0641 & 1.4541 \\ 
              &50& -0.0165 & 1.9941 & -0.0128 & 1.9905 & -0.0287 & 1.9759 & -0.0381 & 1.4775 \\ 
               &60 & -0.0038 & 2.0146 & 0.0295 & 2.0081 & 0.0034 & 2.0220 & -0.0229 & 1.6685 \\ 
  
$(40,30)$&35  & -0.1269 & 1.8712 & -0.0980 & 1.9447 & -0.0898 & 1.9395 & -0.0609 & 1.5198 \\ 
              &50 & -0.0168 & 1.9908 & -0.0128 & 1.9872 & -0.0072 & 2.0051 & -0.0318 & 1.6463 \\ 
              &60 & -0.0064 & 2.0159 & 0.0163 & 2.0240 & 0.0013 & 2.0148 & -0.0195 & 1.8304 \\ 
  
$(50,50)$&70 & -0.0295 & 1.9591 & -0.0333 & 1.9609 & -0.0472 & 1.9395 & -0.0460 & 1.3926 \\ 
              &80 & -0.0121 & 2.0011 & -0.0008 & 2.0031 & -0.0226 & 1.9809 & -0.0305 & 1.5032 \\ 
              &85 & -0.0129 & 1.9987 & 0.0046 & 2.0147 & -0.0210 & 1.9826 & -0.0253 & 1.5771 \\ 
\bottomrule
\end{tabular}
\end{center}
\end{table}
\begin{table}[!ht]
\tabcolsep 1.1pt
\scriptsize
\begin{center}
\caption{\small{Bias and RE values of LS estimates for NIN Bayesian estimates under LINEX loss function.}}
 \label{LS.LINEX.NIN:sem}
\begin{tabular}{cccccccccccccccccccccc}
\toprule
&&&&&&\multicolumn{4}{c}{$v_1=-0.25$}&&&&&&&\multicolumn{4}{c}{$v_2=0.5$}\\
\cmidrule(r){5-12} \cmidrule(r){15-22}
&&&&\multicolumn{2}{c}{${\widehat\theta_1}$}&\multicolumn{2}{c}{${\widehat\theta_2}$}&\multicolumn{2}{c}{${\widehat\theta_3}$}&\multicolumn{2}{c}{${\widehat\theta_4}$}&&&\multicolumn{2}{c}{${\widehat\theta_1}$}&\multicolumn{2}{c}{${\widehat\theta_2}$}&\multicolumn{2}{c}{${\widehat\theta_3}$}&\multicolumn{2}{c}{${\widehat\theta_4}$}\\
\cmidrule(r){5-6} \cmidrule(r){7-8} \cmidrule(r){9-10} \cmidrule(r){11-12}\cmidrule(r){15-16} \cmidrule(r){17-18} \cmidrule(r){19-20} \cmidrule(r){21-22}
$(m,n)$&&$r$&&Bias&RE&Bias&RE&Bias&RE&Bias&RE&&&Bias&RE&Bias&RE&Bias&RE&Bias&RE\\
\hline
$(20,20)$&&20&& 0.0040 & 1.9525 & -0.0299 & 1.9793 & -0.0058 & 1.9688 & -0.0375 & 1.7039 &&& -0.0620 & 2.1643 & -0.0398 & 2.0039 & -0.0640 & 2.0520 & -0.0390 & 1.7037 \\ 
         &&25&& 0.0274 & 1.9901 & 0.0018 & 1.9842 & 0.0190 & 1.9740 & -0.0310 & 1.7364 &&& -0.0170 & 2.0721 & -0.0076 & 2.0410 & -0.0342 & 2.0977 & -0.0326 & 1.7346 \\ 
         &&30& & 0.0420 & 2.0007 & 0.0256 & 2.0010 & 0.0300 & 1.9873 & -0.0195 & 1.8878 &&&  0.0081 & 2.1121 & 0.0166 & 2.0481 & -0.0234 & 2.0620 & -0.0213 & 1.8919 \\ 
  
$(30,40)$&&35& & -0.1425 & 1.9095 & -0.1049 & 1.9652 & -0.1213 & 1.9031 & -0.0639 & 1.4561 &&&  -0.1671 & 1.8895 & -0.1091 & 1.9273 & -0.1402 & 1.8937 & -0.0644 & 1.4504 \\ 
        &&50& & -0.0100 & 1.9838 & -0.0111 & 1.9861 & -0.0216 & 1.9702 & -0.0379 & 1.4778 &&&  -0.0291 & 2.0153 & -0.0161 & 1.9999 & -0.0427 & 1.9887 & -0.0384 & 1.4769 \\
\bottomrule
\end{tabular}
\end{center}
\end{table}
\begin{table}[!ht]
\tabcolsep 1.1pt
\scriptsize
\begin{center}
\caption{\small{Bias and RE values of LS estimates for NIN Bayesian estimates under GE loss function.}}
 \label{LS.GE.NIN:sem}
\begin{tabular}{cccccccccccccccccccccc}
\toprule
&&&&&&\multicolumn{4}{c}{$k_1=-0.25$}&&&&&&&\multicolumn{4}{c}{$k_2=0.5$}\\
\cmidrule(r){5-12} \cmidrule(r){15-22}
&&&&\multicolumn{2}{c}{${\widehat\theta_1}$}&\multicolumn{2}{c}{${\widehat\theta_2}$}&\multicolumn{2}{c}{${\widehat\theta_3}$}&\multicolumn{2}{c}{${\widehat\theta_4}$}&&&\multicolumn{2}{c}{${\widehat\theta_1}$}&\multicolumn{2}{c}{${\widehat\theta_2}$}&\multicolumn{2}{c}{${\widehat\theta_3}$}&\multicolumn{2}{c}{${\widehat\theta_4}$}\\
\cmidrule(r){5-6} \cmidrule(r){7-8} \cmidrule(r){9-10} \cmidrule(r){11-12}\cmidrule(r){15-16} \cmidrule(r){17-18} \cmidrule(r){19-20} \cmidrule(r){21-22}
$(m,n)$&&$r$&&Bias&RE&Bias&RE&Bias&RE&Bias&RE&&&Bias&RE&Bias&RE&Bias&RE&Bias&RE\\
\hline

$(20,20)$&&20& & -0.0595 & 2.1283 & -0.0439 & 2.1325 & -0.0544 & 2.0472 & -0.0411 & 1.6837 &&& -0.1012 & 2.0906 & -0.0555 & 2.0998 & -0.0839 & 2.0285 & -0.0445 & 1.6622 \\ 
           &&25&& -0.0146 & 1.9943 & -0.0104 & 2.0106 & -0.0250 & 1.9665 & -0.0347 & 1.6540 &&& -0.0417 & 1.9907 & -0.0200 & 2.0059 & -0.0510 & 1.9666 & -0.0380 & 1.6379 \\ 
         &&30& & 0.0104 & 1.9208 & 0.0143 & 1.9583 & -0.0135 & 1.9877 & -0.0235 & 1.7098 &&& -0.0099 & 1.9344 & 0.0058 & 1.9709 & -0.0387 & 1.9862 & -0.0270 & 1.7014 \\ 
  
$(30,40)$&&35& & -0.1700 & 2.2310 & -0.1117 & 2.2636 & -0.1383 & 2.0195 & -0.0651 & 1.5548 &&& -0.1896 & 2.1434 & -0.1174 & 2.1877 & -0.1492 & 1.9865 & -0.0661 & 1.5325 \\ 
       &&50&& -0.0288 & 1.9660 & -0.0177 & 2.0275 & -0.0394 & 1.9522 & -0.0392 & 1.4535 &&& -0.0413 & 1.9561 & -0.0228 & 2.0145 & -0.0501 & 1.9442 & -0.0405 & 1.4468 \\
\bottomrule
\end{tabular}
\end{center}
\end{table}
\begin{table}[!ht]
\tabcolsep 13pt
\scriptsize
\begin{center}
\caption{\small{Bias and RE values of SP estimates for EM estimations.}}
 \label{SPT.EM:sim}

\begin{tabular}{cccccccccc}
\toprule
&&\multicolumn{2}{c}{${\widehat\theta_1}$}&\multicolumn{2}{c}{${\widehat\theta_2}$}&\multicolumn{2}{c}{${\widehat\theta_3}$}&\multicolumn{2}{c}{${\widehat\theta_4}$}\\
\cmidrule(lr){3-4} \cmidrule(lr){5-6} \cmidrule(lr){7-8} \cmidrule(lr){9-10}
$(m,n)$&$r$&Bias&RE&Bias&RE&Bias&RE&Bias&RE\\
  \hline
$(20,20)$&20& 0.4945 & 1.0036 & 0.1666 & 1.0050 & 0.2519 & 1.0046 & 0.0360 & 1.0013 \\ 
              &25& 0.2819 & 1.0023 & 0.1205 & 1.0030 & 0.2437 & 1.0022 & 0.0349 & 1.0017 \\ 
              &30& 0.1556 & 1.0020 & 0.0987 & 1.0020 & 0.1720 & 1.0024 & 0.0396 & 1.0003 \\ 
  
$(30,40)$&35&  0.2919 & 1.0019 & 0.0908 & 1.0033 & 0.1524 & 1.0021 & 0.0223 & 1.0007 \\ 
              &50& 0.1012 & 1.0012 & 0.0617 & 1.0017 & 0.0806 & 1.0015 & 0.0187 & 0.9996 \\ 
              &60 & 0.0371 & 1.0010 & 0.0570 & 1.0011 & 0.0709 & 1.0011 & 0.0125 & 0.9997 \\ 
  
$(40,30)$&35&  0.1821 & 1.0017 & 0.0764 & 1.0033 & 0.1784 & 1.0020 & 0.0249 & 1.0013 \\ 
              &50 & 0.0752 & 1.0011 & 0.0446 & 1.0016 & 0.1130 & 1.0012 & 0.0169 & 1.0002 \\ 
              &60 & 0.0412 & 1.0012 & 0.0328 & 1.0014 & 0.0913 & 1.0011 & 0.0169 & 0.9998 \\ 
  
$(50,50)$&70& 0.0630 & 1.0006 & 0.0406 & 1.0010 & 0.0602 & 1.0006 & 0.0158 & 0.9994 \\
              &80 & 0.0421 & 1.0006 & 0.0306 & 1.0009 & 0.0557 & 1.0008 & 0.0136 & 0.9996 \\ 
              &85 & 0.0264 & 1.0011 & 0.0270 & 1.0008 & 0.0439 & 1.0005 & 0.0088 & 0.9998 \\ 

\bottomrule
\end{tabular}
\end{center}
\end{table}

\begin{table}[!ht]
\tabcolsep 12.5pt
\scriptsize
\begin{center}
\caption{\small{Bias and RE values of SP estimates for IN Bayesian estimates under SE loss function.}}
 \label{SPT.SEL.IN:sim}
\begin{tabular}{cccccccccc}
\toprule
&&\multicolumn{2}{c}{${\widehat\theta_1}$}&\multicolumn{2}{c}{${\widehat\theta_2}$}&\multicolumn{2}{c}{${\widehat\theta_3}$}&\multicolumn{2}{c}{${\widehat\theta_4}$}\\
\cmidrule(lr){3-4} \cmidrule(lr){5-6} \cmidrule(lr){7-8} \cmidrule(lr){9-10}
$(m,n)$&$r$&Bias&RE&Bias&RE&Bias&RE&Bias&RE\\
\hline
$(20,20)$&20& -0.0278 & 1.0637 & -0.0372 & 1.0278 & 0.0007 & 1.0171 & -0.0144 & 1.0066 \\ 
              &25 &  0.0228 & 1.0132 & 0.0130 & 1.0098 & 0.0351 & 1.0093 & -0.0020 & 1.0038 \\  
              &30 & 0.0656 & 1.0049 & 0.0428 & 1.0056 & 0.0582 & 1.0063 & 0.0127 & 1.0014 \\  
  
$(30,40)$&35& -0.1412 & 1.0166 & -0.1390 & 1.0047 & -0.1479 & 1.0053 & -0.0672 & 1.0012 \\ 
               &50 & 0.0313 & 1.0023 & -0.0098 & 1.0018 & -0.0002 & 1.0021 & -0.0197 & 1.0013 \\
              &60 & 0.0380 & 1.0009 & 0.0352 & 1.0015 & 0.0593 & 1.0013 & 0.0065 & 0.9993 \\ 
  
$(40,30)$&35& -0.1246 & 1.0096 & -0.1366 & 1.0034 & -0.1208 & 1.0047 & -0.0578 & 1.0031 \\  
               &50& 0.0161 & 1.0019 & -0.0111 & 1.0023 & 0.0140 & 1.0022 & -0.0127 & 1.0003 \\ 
               &60 & 0.0181 & 1.0018 & 0.0241 & 1.0011 & 0.0477 & 1.0016 & 0.0105 & 0.9997 \\  
  
 $(50,50)$&70& -0.0111 & 1.0015 & -0.0532 & 1.0011 & -0.0563 & 1.0012 & -0.0349 & 1.0000 \\  
               &80& 0.0130 & 1.0011 & 0.0051 & 1.0009 & -0.0035 & 1.0012 & -0.0042 & 0.9998 \\ 
               &85& 0.0251 & 1.0007 & 0.0231 & 1.0010 & 0.0309 & 1.0011 & 0.0017 & 1.0004 \\ 
\bottomrule
\end{tabular}
\end{center}
\end{table}
\begin{table}[!ht]
\tabcolsep 1.1pt
\scriptsize
\begin{center}
\caption{\small{Bias and RE values of SP estimates for IN Bayesian estimates under LINEX loss function.}}
 \label{SPT.LINEX.IN:sem}
\begin{tabular}{cccccccccccccccccccccc}
\toprule
&&&&&&\multicolumn{4}{c}{$v_1=-0.25$}&&&&&&&&\multicolumn{4}{c}{$v_2=0.5$}\\
\cmidrule(r){5-12} \cmidrule(r){15-22}
&&&&\multicolumn{2}{c}{${\widehat\theta_1}$}&\multicolumn{2}{c}{${\widehat\theta_2}$}&\multicolumn{2}{c}{${\widehat\theta_3}$}&\multicolumn{2}{c}{${\widehat\theta_4}$}&&&\multicolumn{2}{c}{${\widehat\theta_1}$}&\multicolumn{2}{c}{${\widehat\theta_2}$}&\multicolumn{2}{c}{${\widehat\theta_3}$}&\multicolumn{2}{c}{${\widehat\theta_4}$}\\
\cmidrule(r){5-6} \cmidrule(r){7-8} \cmidrule(r){9-10} \cmidrule(r){11-12}\cmidrule(r){15-16} \cmidrule(r){17-18} \cmidrule(r){19-20} \cmidrule(r){21-22}
$(m,n)$&&$r$&&Bias&RE&Bias&RE&Bias&RE&Bias&RE&&&Bias&RE&Bias&RE&Bias&RE&Bias&RE\\
\hline

$(20,20)$&&20& & -0.0042 & 1.0624 & -0.0321 & 1.0273 & 0.0360 & 1.0159 & -0.0134 & 1.0073 &&& -0.0611 & 1.0051 & -0.0462 & 1.0012 & -0.0596 & 1.0019 & -0.0162 & 1.0067 \\ 
        &&25&& 0.0446 & 1.0126 & 0.0185 & 1.0094 & 0.0661 & 1.0087 & -0.0010 & 1.0036  &&& -0.0140 & 1.0023 & 0.0030 & 1.0006 & -0.0228 & 1.0011 & -0.0040 & 1.0039 \\ 
      &&30& & 0.0831 & 1.0043 & 0.0480 & 1.0054 & 0.0885 & 1.0059 & 0.0138 & 1.0012  &&& 0.0329 & 1.0016 & 0.0330 & 1.0007 & 0.0005 & 1.0007 & 0.0106 & 1.0013 \\ 
  
$(30,40)$&&35& & -0.1276 & 1.0192 & -0.1363 & 1.0048 & -0.1354 & 1.0050 & -0.0669 & 1.0013  &&& -0.1665 & 1.0010 & -0.1446 & 1.0001 & -0.1727 & 1.0019 & -0.0677 & 1.0014 \\ 
           &&50& & 0.0425 & 1.0024 & -0.0068 & 1.0019 & 0.0134 & 1.0024 & -0.0194 & 1.0015  &&& 0.0098 & 1.0006 & -0.0158 & 1.0001 & -0.0269 & 1.0004 & -0.0205 & 1.0012 \\
\bottomrule
\end{tabular}
\end{center}
\end{table}

\begin{table}[!ht]
\tabcolsep 1.1 pt
\scriptsize
\begin{center}
\caption{\small{Bias and RE values of SP estimates for IN Bayesian estimates under GE loss function.}}
 \label{SPT.GEL.IN:sem}
\begin{tabular}{cccccccccccccccccccccc}
\toprule
&&&&&&\multicolumn{4}{c}{$k_1=-0.25$}&&&&&&&\multicolumn{4}{c}{$k_2=0.5$}\\
\cmidrule(r){5-12} \cmidrule(r){15-22}
&&&&\multicolumn{2}{c}{${\widehat\theta_1}$}&\multicolumn{2}{c}{${\widehat\theta_2}$}&\multicolumn{2}{c}{${\widehat\theta_3}$}&\multicolumn{2}{c}{${\widehat\theta_4}$}&&&\multicolumn{2}{c}{${\widehat\theta_1}$}&\multicolumn{2}{c}{${\widehat\theta_2}$}&\multicolumn{2}{c}{${\widehat\theta_3}$}&\multicolumn{2}{c}{${\widehat\theta_4}$}\\
\cmidrule(r){5-6} \cmidrule(r){7-8} \cmidrule(r){9-10} \cmidrule(r){11-12}\cmidrule(r){15-16} \cmidrule(r){17-18} \cmidrule(r){19-20} \cmidrule(r){21-22}
$(m,n)$&&$r$&&Bias&RE&Bias&RE&Bias&RE&Bias&RE&&&Bias&RE&Bias&RE&Bias&RE&Bias&RE\\
\hline
$(20,20)$ &&20&& -0.0740 & 1.0442 & -0.0531 & 1.0204 & -0.0500 & 1.0145 & -0.0202 & 1.0089 &&& -0.1218 & 1.0385 & -0.0702 & 1.0187 & -0.1002 & 1.0138 & -0.0262 & 1.0092 \\
 &&25&& -0.0168 & 1.0147 & -0.0030 & 1.0101 & -0.0103 & 1.0111 & -0.0079 & 1.0057 &&& -0.0571 & 1.0140 & -0.0197 & 1.0085 & -0.0574 & 1.0112 & -0.0143 & 1.0044 \\ 
 &&30& & 0.0329 & 1.0062 & 0.0283 & 1.0069 & 0.0153 & 1.0070 & 0.0066 & 1.0028 &&& 0.0005 & 1.0064 & 0.0137 & 1.0067 & -0.0280 & 1.0070 & 0.0004 & 1.0042 \\ 
 
$(30,40)$ &&35&& -0.1719 & 1.0097 & -0.1482 & 1.0030 & -0.1678 & 1.0048 & -0.0689 & 1.0018 &&& -0.2039 & 1.0094 & -0.1576 & 1.0030 & -0.1886 & 1.0041 & -0.0707 & 1.0020 \\
  &&50& & 0.0100 & 1.0031 & -0.0190 & 1.0022 & -0.0217 & 1.0025 & -0.0221 & 1.0012 &&& -0.0119 & 1.0034 & -0.0284 & 1.0019 & -0.0422 & 1.0022 & -0.0246 & 1.0011 \\
\bottomrule
\end{tabular}
\end{center}
\end{table}
\begin{table}[!ht]
\tabcolsep 12.5pt
\scriptsize
\begin{center}
\caption{\small{Bias and RE values of SP estimates for NIN Bayesian estimates under SE loss function.}}
 \label{SPT.SEL.NON:sim}
\begin{tabular}{cccccccccc}
\toprule
&&\multicolumn{2}{c}{${\widehat\theta_1}$}&\multicolumn{2}{c}{${\widehat\theta_2}$}&\multicolumn{2}{c}{${\widehat\theta_3}$}&\multicolumn{2}{c}{${\widehat\theta_4}$}\\
\cmidrule(lr){3-4} \cmidrule(lr){5-6} \cmidrule(lr){7-8} \cmidrule(lr){9-10}
$(m,n)$&$r$&Bias&RE&Bias&RE&Bias&RE&Bias&RE\\
\hline
$(20,20)$&20 & -0.0130 & 1.0558 & -0.0594 & 1.0189 & -0.0055 & 1.0099 & -0.0267 & 1.0048 \\ 
              &25 & 0.0696 & 1.0097 & 0.0068 & 1.0056 & 0.0484 & 1.0061 & -0.0136 & 1.0029 \\ 
              &30& 0.1084 & 1.0024 & 0.0540 & 1.0027 & 0.0710 & 1.0035 & 0.0093 & 1.0010 \\ 
  
$(30,40)$&35 & -0.2545 & 1.0100 & -0.2021 & 1.0023 & -0.2060 & 1.0032 & -0.0781 & 1.0019 \\
              &50& 0.0166 & 1.0017 & -0.0154 & 1.0014 & -0.0069 & 1.0030 & -0.0262 & 1.0010 \\
              &60& 0.0424 & 1.0009 & 0.0689 & 1.0002 & 0.0574 & 1.0016 & 0.0042 & 1.0002 \\ 
  
$(40,30)$&35 & -0.2033 & 1.0047 & -0.1854 & 1.0015 & -0.1285 & 1.0041 & -0.0716 & 1.0025 \\ 
              &50& 0.0162 & 1.0017 & -0.0156 & 1.0016 & 0.0352 & 1.0018 & -0.0136 & 1.0015 \\
              &60 & 0.0372 & 1.0012 & 0.0424 & 1.0009 & 0.0525 & 1.0011 & 0.0108 & 1.0001 \\ 
  
$(50,50)$& 70& -0.0090 & 1.0012 & -0.0567 & 1.0007 & -0.0442 & 1.0015 & -0.0421 & 1.0003 \\
                &80 & 0.0256 & 1.0006 & 0.0081 & 1.0007 & 0.0046 & 1.0007 & -0.0110 & 1.0000 \\
               &85 & 0.0240 & 1.0006 & 0.0191 & 1.0006 & 0.0078 & 1.0007 & -0.0005 & 1.0001 \\ 
\bottomrule
\end{tabular}
\end{center}
\end{table}
\begin{table}[!ht]
\tabcolsep 1.1pt
\scriptsize
\begin{center}
\caption{\small{Bias and RE values of SP estimates for NIN Bayesian estimates under LINEX loss function.}}
 \label{SPT.LINEX.NON:sem}
\begin{tabular}{cccccccccccccccccccccc}
\toprule
&&&&&&\multicolumn{4}{c}{$v_1=-0.25$}&&&&&&&\multicolumn{4}{c}{$v_2=0.5$}\\
\cmidrule(r){5-12} \cmidrule(r){15-22}
&&&&\multicolumn{2}{c}{${\hat\theta_1}$}&\multicolumn{2}{c}{${\widehat\theta_2}$}&\multicolumn{2}{c}{${\widehat\theta_3}$}&\multicolumn{2}{c}{${\widehat\theta_4}$}&&&\multicolumn{2}{c}{${\widehat\theta_1}$}&\multicolumn{2}{c}{${\widehat\theta_2}$}&\multicolumn{2}{c}{${\widehat\theta_3}$}&\multicolumn{2}{c}{${\widehat\theta_4}$}\\
\cmidrule(r){5-6} \cmidrule(r){7-8} \cmidrule(r){9-10} \cmidrule(r){11-12}\cmidrule(r){15-16} \cmidrule(r){17-18} \cmidrule(r){19-20} \cmidrule(r){21-22}
$(m,n)$&&$r$&&Bias&RE&Bias&RE&Bias&RE&Bias&RE&&&Bias&RE&Bias&RE&Bias&RE&Bias&RE\\
\hline
$(20,20)$&&20 && 0.0227 & 1.0905 & -0.0527 & 1.0173 & 0.0338 & 1.0085 & -0.0257 & 1.0051 &&& -0.0750 & 1.0041 & -0.0695 & 1.0005 & -0.0775 & 1.0044 & -0.0287 & 1.0046 \\ 
        &&25&& 0.0994 & 1.0154 & 0.0133 & 1.0057 & 0.0851 & 1.0061 & -0.0125 & 1.0033 &&& 0.0165 & 1.0012 & -0.0051 & 1.0004 & -0.0182 & 1.0007 & -0.0157 & 1.0030 \\ 
           &&30& & 0.1327 & 1.0023 & 0.0601 & 1.0030 & 0.1078 & 1.0032 & 0.0105 & 1.0010 &&& 0.0664 & 1.0006 & 0.0431 & 1.0005 & 0.0033 & 1.0003 & 0.0069 & 1.0010 \\ 
           
$(30,40)$&&35 && -0.2386 & 1.0125 & -0.1995 & 1.0023 & -0.1935 & 1.0032 & -0.0779 & 1.0019 &&& -0.2844 & 1.0003 & -0.2082 & 1.0001 & -0.2303 & 1.0004 & -0.0787 & 1.0019 \\ 
           &&50&& 0.0297 & 1.0016 & -0.0122 & 1.0014 & 0.0072 & 1.0027 & -0.0258 & 1.0008 &&& -0.0082 & 1.0002 & -0.0221 & 1.0001 & -0.0353 & 1.0004 & -0.0269 & 1.0010 \\
\bottomrule
\end{tabular}
\end{center}
\end{table}

\begin{table}[!ht]
\tabcolsep 1.1pt
\scriptsize
\begin{center}
\caption{\small{Bias and RE values of SP estimates for NIN Bayesian estimates under GE loss function.}}
 \label{SPT.GEL.NON:sem}
\begin{tabular}{cccccccccccccccccccccc}
\toprule
&&&&&&\multicolumn{4}{c}{$k_1=-0.25$}&&&&&&&\multicolumn{4}{c}{$k_2=0.5$}\\
\cmidrule(r){5-12} \cmidrule(r){15-22}
&&&&\multicolumn{2}{c}{${\widehat\theta_1}$}&\multicolumn{2}{c}{${\widehat\theta_2}$}&\multicolumn{2}{c}{${\widehat\theta_3}$}&\multicolumn{2}{c}{${\widehat\theta_4}$}&&&\multicolumn{2}{c}{${\widehat\theta_1}$}&\multicolumn{2}{c}{${\widehat\theta_2}$}&\multicolumn{2}{c}{${\widehat\theta_3}$}&\multicolumn{2}{c}{${\widehat\theta_4}$}\\
\cmidrule(r){5-6} \cmidrule(r){7-8} \cmidrule(r){9-10} \cmidrule(r){11-12}\cmidrule(r){15-16} \cmidrule(r){17-18} \cmidrule(r){19-20} \cmidrule(r){21-22}
$(m,n)$&&$r$&&Bias&RE&Bias&RE&Bias&RE&Bias&RE&&&Bias&RE&Bias&RE&Bias&RE&Bias&RE\\
\hline
$(20,20)$&&20&& -0.0893 & 1.0200 & -0.0796 & 1.0100 & -0.0624 & 1.0100 & -0.0332 & 1.0050 &&& -0.1669 & 1.0175 & -0.1030 & 1.0098 & -0.1210 & 1.0083 & -0.0398 & 1.0054 \\
        &&25& & 0.0177 & 1.0066 & -0.0113 & 1.0055 & -0.0018 & 1.0069 & -0.0198 & 1.0053 &&& -0.0376 & 1.0064 & -0.0306 & 1.0053 & -0.0542 & 1.0064 & -0.0263 & 1.0055 \\ 
           &&30& & 0.0680 & 1.0037 & 0.0374 & 1.0040 & 0.0220 & 1.0039 & 0.0025 & 1.0021 &&& 0.0285 & 1.0037 & 0.0211 & 1.0042 & -0.0277 & 1.0042 & -0.0046 & 1.0012 \\ 
  
 $(30,40)$&&35&& -0.2901 & 1.0025 & -0.2128 & 1.0010 & -0.2268 & 1.0024 & -0.0801 & 1.0017 &&& -0.3315 & 1.0058 & -0.2238 & 1.0010 & -0.2477 & 1.0024 & -0.0821 & 1.0022 \\ 
           &&50&& -0.0082 & 1.0018 & -0.0251 & 1.0017 & -0.0286 & 1.0029 & -0.0285 & 1.0015 &&& -0.0334 & 1.0015 & -0.0352 & 1.0016 & -0.0499 & 1.0026 & -0.0309 & 1.0021 \\
\bottomrule
\end{tabular}
\end{center}
\end{table}
\begin{table}[!ht]
\tabcolsep 3.5pt
\scriptsize
\begin{center}
\caption{\small{95\% L and U bound estimates and ILs  of EM and Bayesian estimations}.}
\label{L.U.CP:sim}
\begin{tabular}{cccccccccccccccccccc}
\toprule
&&&&\multicolumn{3}{c}{${\widehat\theta_1}$}&&\multicolumn{3}{c}{${\widehat\theta_2}$}&&\multicolumn{3}{c}{${\widehat\theta_3}$}&&\multicolumn{3}{c}{${\widehat\theta_4}$}\\
\cmidrule{5-7}\cmidrule{9-11}\cmidrule{13-15}\cmidrule{17-19}

$(m,n)$&r&&&L&U&IL&&L&U&IL&&L&U&IL&&L&U&IL\\
\hline
$(20,20)$&20&&ACI& 0.1165 & 4.2485 & 4.1320 && 0.4549 & 1.8846 & 1.4297 && 0.7493 & 3.7663 & 3.0171 && 0.2913 & 0.7832 & 0.4919 \\
           &&NIN&CrI&0.6145 & 3.1811 & 2.5666 &&0.5274 & 1.6047 & 1.0772 && 1.0417 & 3.5985 & 2.5568 && 0.2936 & 0.7215 & 0.4278 \\ 
             &&&HPD&0.5684 & 2.9384 & 2.3700 && 0.4765 & 1.5107 & 1.0341&& 0.8836 & 3.2799 & 2.3964 && 0.2709 & 0.6861 & 0.4152 \\ 
           &&IN&CrI&0.8349 & 2.4081 & 1.5731 && 0.5842 & 1.4143 & 0.8302 && 1.1355 & 3.1633 & 2.0278 && 0.3249 & 0.7119 & 0.3870 \\ 
     &&&HPD&0.7538 & 2.2554 & 1.5016 && 0.5807 & 1.3956 & 0.8149 && 1.0618 & 2.9969 & 1.9351 && 0.3129 & 0.6750 & 0.3621 \\ [2pt]

&25&&ACI& 0.5003 & 3.0754 & 2.5751 && 0.5633 & 1.6794 & 1.1162&& 0.9417 & 3.5533 & 2.6116 && 0.3108 & 0.7599 & 0.4491 \\
\bottomrule
\end{tabular}
\end{center}
\end{table}
\begin{table}[!ht]
\tabcolsep 2pt
\scriptsize
\begin{center}
\caption{\small{Predictive estimates of $w_t=w_{r+j}$ for $j=1,2$.}}
\label{P.E:sim}
\begin{tabular}{cccccccccccccccc}
\toprule
&&&\multicolumn{1}{c}{Classical prediction}&&\multicolumn{3}{c}{NIN Bayesian prediction}&&&&\multicolumn{3}{c}{IN Bayesian prediction}\\
\cmidrule(r){4-4}\cmidrule(r){5-9}\cmidrule(r){11-15}
&&&\multicolumn{1}{c}{BUP}&\multicolumn{1}{c}{SE}&\multicolumn{2}{c}{LINEX}&\multicolumn{2}{c}{GE}&&\multicolumn{1}{c}{SE}&\multicolumn{2}{c}{LINEX}&\multicolumn{2}{c}{GE}\\
\cmidrule(r){6-7}\cmidrule(r){8-9}\cmidrule(r){12-13}\cmidrule(r){14-15}
$(m,n)$&r&j&&&$v_1$=-0.25&$v_2$=0.5&$k_1=-0.25$&$k_2=0.5$&&&$v_1$=-0.25&$v_2$=0.5&$k_1=-0.25$&$k_2=0.5$\\
\hline
$(20,20)$&20&1& 0.4284 & 0.4487 & 0.4375 & 0.4338 & 0.4219 & 0.4211 && 0.4467 & 0.4450 & 0.4397 & 0.4270 & 0.4112 \\ 
  &&2 & 0.4826 & 0.5181 & 0.5047 & 0.5043 & 0.4925 & 0.5138 && 0.5100 & 0.5071 & 0.5039 & 0.4914 & 0.4773 \\ 
  
 &25&1 &  0.7671 & 0.7921 & 0.8045 & 0.7796 & 0.7655 & 0.7557 && 0.8032 & 0.8077 & 0.7834 & 0.7678 & 0.7371 \\ 
 &&2 &  0.8684 & 0.9097 & 0.9263 & 0.8974 & 0.8853 & 0.9022 && 0.9165 & 0.9216 & 0.8956 & 0.8818 & 0.8519 \\ 
  
 &30&1 & 1.5040 & 1.5504 & 1.5905 & 1.4646 & 1.4706 & 1.4351 && 1.5407 & 1.5621 & 1.4335 & 1.4376 & 1.3720 \\
\bottomrule              
\end{tabular}
\end{center}
\end{table}
\begin{table}[!ht]
\tabcolsep 1.85pt
\scriptsize
\begin{center}
\caption{Predictive interval estimates of $w_t=w_{r+j}$ for $j=1,2$.}
\label{P.I.E:sim}
\begin{tabular}{cccccccccccccccccccccccc}
\toprule
&&&&&\multicolumn{3}{c}{Classical prediction}&&\multicolumn{7}{c}{NIN Bayesian prediction}&&\multicolumn{7}{c}{IN Bayesian prediction}\\
\cmidrule(r){6-8}\cmidrule(r){10-16}\cmidrule(r){18-24}
&&&&&&&&&\multicolumn{3}{c}{CrI}&&\multicolumn{3}{c}{HPD}&&\multicolumn{3}{c}{CrI}&&\multicolumn{3}{c}{HPD}\\
\cmidrule(r){6-8}\cmidrule(r){10-12}\cmidrule(r){14-16}\cmidrule(r){18-20}\cmidrule(r){22-24}
$(m,n)$&r&&j&&$L_{w_{t}}$&$U_{w_{t}}$&\text{IL}&&$L_{w_{t}}$&$U_{w_{t}}$&\text{IL}&&$L_{w_{t}}$&$U_{w_{t}}$&\text{IL}&&$L_{w_{t}}$&$U_{w_{t}}$&\text{IL}&&$L_{w_{t}}$&$U_{w_{t}}$&\text{IL}\\
\hline

$(20,20)$&20&&1&& 0.1848 & 0.8288 & 0.6440 && 0.2012 & 0.8806 & 0.6794 && 0.1828 & 0.8188 & 0.6360 && 0.1913 & 0.8229 & 0.6315 && 0.1775 & 0.7950 & 0.6174 \\ [1pt]
&&&2&&0.2074 & 0.9628 & 0.7554 && 0.2403 & 1.0107 & 0.7704 &&  0.1887 & 0.9189 & 0.7302 &&  0.2294 & 0.9292 & 0.6998 &&  0.1942 & 0.8847 & 0.6905 \\ [1pt]
&25&&1&&0.3849 & 1.4968 & 1.1119 &&  0.3703 & 1.5073 & 1.1369 &&  0.3132 & 1.3572 & 1.0440 &&  0.3649 & 1.5573 & 1.1924&& 0.2877 & 1.3664 & 1.0787 \\ [1pt]
&&&2&&0.4238 & 1.7290 & 1.3053 &&  0.4238 & 1.7963 & 1.3724 &&  0.3594 & 1.5826 & 1.2232 &&  0.4255 & 1.7512 & 1.3257 && 0.3354 & 1.5846 & 1.2493 \\ [1pt]
&30&&1&&0.7103 & 3.0056 & 2.2953 &&  0.6735 & 3.1763 & 2.5028 &&  0.5846 & 2.9248 & 2.3403 &&  0.6889 & 3.1254 & 2.4365 &&  0.5585 & 2.8765 & 2.3180 \\
\bottomrule                           
\end{tabular} 
\end{center}
\end{table}

\AtVeryEndDocument {Empty hook}

\end{document}